\documentclass[a4paper,aps,pre,superscriptaddress,floatfix,nofootinbib,longbibliography,notitlepage,twocolumn]{revtex4-2}
\usepackage{ragged2e}
\usepackage{bbm}
\usepackage{bbold}
\usepackage[pdftex]{graphicx}
\usepackage{latexsym,amsmath,verbatim,amssymb,txfonts}
\usepackage{lipsum}
\usepackage{hyperref}
\usepackage{color}
\usepackage{svg}
\usepackage{rotating}
\usepackage{verbatim}
\usepackage{multirow}
\usepackage[english]{babel}
\usepackage{comment}
\usepackage{bm}
\usepackage[utf8]{inputenc}
\usepackage{enumerate}
\usepackage{tikz}
\usepackage{color}
\usepackage{braket}
\usepackage{pifont}
\usepackage{blkarray, bigstrut}
\usepackage{xcolor}
\usepackage{graphicx}
\usepackage{subcaption}
\usepackage{orcidlink}
\usepackage[normalem]{ulem}
\newcommand{\Th}[1]{\textcolor{blue}{#1}}

\begin{document}
\title{Synchronization of coupled wind turbines}
 
\author{Nadia Kevine Kouonang}
\affiliation{Department of Mathematics \& naXys, Namur Institute for Complex Systems, University of Namur, B5000 Namur, Belgium}
\affiliation{Mechanics and Modeling of Physical Systems Research Unit (UR-2MSP), Department of Physics, Faculty of Sciences, University of Dschang, Dschang BP 067, Cameroon 
}

\author{Jeanne Sandrine Takam Mabekou}
\affiliation{Mechanics and Modeling of Physical Systems Research Unit (UR-2MSP), Department of Physics, Faculty of Sciences, University of Dschang, Dschang BP 067, Cameroon 
}

\author{Thierry Njougouo\,\orcidlink{0000-0001-7706-7674}}
\affiliation{IMT School for Advanced Studies, Piazza San Francesco 19, 55100 Lucca, Italy}
 
\author{Timoteo Carletti\,\orcidlink{0000-0003-2596-4503}}
\affiliation{Department of Mathematics \& naXys, Namur Institute for Complex Systems, University of Namur, B5000 Namur, Belgium}

\begin{abstract}
In the context of renewable energies, wind energy appears as a sustainable alternative to address current environmental and energy challenges. This work studies the synchronization and stability of a network of wind turbines  subjected to strong disturbances, by integrating a realistic modeling of wind variability by using the Ornstein-Uhlenbeck stochastic process. The dynamics of each wind turbine are described by a Kuramoto-type equation, while synchronization is analyzed through the time evolution of the phases. Stability is studied by analyzing the basin of attraction to the synchronous solution, namely the set of initial conditions leading to the stable synchronous state. Simulations carried out on various models ranging from an isolated wind turbine with constant power to an isolated wind turbine with variable wind power, reveal that the stability of the system is strongly influenced by inertia, damping, wind speed, wind fluctuation rate, correlation time, and coupling strength.
{Physically, these parameters control the balance between injected mechanical power, energy dissipation, grid-induced restoring forces, and the temporal structure of wind fluctuations, thereby determining the ability of the wind turbine to absorb perturbations and maintain synchronization under fluctuating wind conditions.}
\end{abstract}

\maketitle

\section{Introduction}
\label{sec:intro}
The world population is experiencing continuous growth that together with the increase of human activities such as, agriculture, industry and technology, lead to an rise in energy demand~\cite{efstathios2012alternative,bhandari2014novel,anagnostopoulos2007pumping,black2006value}. Faced to this energy pressure, the use of fossil resources such as coal, oil, or natural gas present several constraints, namely: a progressive depletion over time, an exploitation that becomes increasingly costly, and a use that strongly contributes to greenhouse gas emissions in the atmosphere, thereby aggravating climate change~\cite{luna2012optimal,chauhan2014review}. These observations have generated a growing interest in renewable energies, perceived as a sustainable solution, less polluting and better adapted to long-term energy challenges~\cite{carter1951analyzing,nguyen2020electric}. Among the renewable energy sources, wind energy stands out due to its ability to convert the kinetic energy of the wind into clean electricity thanks to the turbines~\cite{luna2012optimal}. 

Under normal operating regime, a power network evolves in a synchronous state, where all frequencies, i.e., angular velocities, are identical to the nominal one, i.e., $50~\mathrm{Hz}$ or $60~\mathrm{Hz}$~\cite{filatrella2008analysis}. In this state, stationary power flows make it possible to maintain a permanent balance between the production and the consumption of energy at all the nodes of the network. When certain parts of the network lose synchronization, potentially destructive frequency oscillations may appear and damage the equipment. In order to limit these damages, the affected components must be isolated from the rest of the network. However, these local disconnections may lead to the desynchronization of other parts of the system, thereby triggering a succession of outages that may lead to large-scale failures. Recent examples of such events are reported in~\cite{handbook2004union,central2012report}, while detailed analyses of large-scale failures are presented in~\cite{motter2002cascade,schafer2018dynamically}. The failure of a transmission line during a blackout does not depend only on the topological structure of the network or on the static distribution of electrical flows. It is also influenced by the collective transient dynamics of the system, whose instabilities may appear on time scales of the order of the second~\cite{schafer2018non}. In general, power networks are designed in such a way as to make the synchronous state locally stable, in order to prevent a simple small-amplitude perturbation from triggering a cascading desynchronization. However, even if this synchronous state remains robust to small fluctuations, the state space of a power network may also contain several stable non-synchronous states toward which the system may switch following a short-circuit, due to variations in renewable energy production, or other significant perturbations, e.g., changes in wind velocity~\cite{schafer2018non,chiang2011direct,tyloo2019noise}. 

The objective of this work is to study the synchronization and the global stability of a network of wind turbines subjected to significant perturbations, while giving particular attention to the realistic modeling of the temporal variability of the wind; the latter is described by using the Ornstein-Uhlenbeck process~\cite{hu2010parameter} , which makes it possible to represent more faithfully the fluctuations observed in real conditions~\cite{luna2012optimal}. The dynamics of each wind generator are modeled by using a Kuramoto model with inertia~\cite{batista2021secondary,filatrella2008analysis,tchuisseu2023secondary,acebron2005kuramoto}. The analysis of synchronization then relies on the study of the temporal evolution of the phases, the associated velocities, i.e., the frequencies, as well as on the evaluation of the order parameter and frequency variation. Synchronization constitutes an essential condition for the correct functioning of a wind turbines network~\cite{filatrella2008analysis,dorfler2014synchronization}. It is therefore of fundamental interest to explore the link between the properties of the network and its stability in the presence of large perturbations~\cite{schafer2018non,schafer2015decentral,menck2013basin}. To achieve this goal we resort to the study of basin of stability of the synchronous solution, namely the set of initial conditions whose solution converges toward the synchronous state. This analysis has been performed under fixed wind condition as well as with stochastic wind variability. The obtained results show that the size of the basin of stability strongly depends on the model parameters: inertia, damping, wind speed, the fluctuation rate of the wind speed, the correlation time, and the coupling strength. By understanding the impact of the structural parameters of the wind turbine, will allow to propose new designs for which the basin of stability is larger and thus the system be able to absorb perturbations and return to the synchronous state.

The paper is organized as follows: Section~\ref{sec:model} presents the proposed model of the wind turbine network, where wind speed fluctuations are described by an Ornstein–Uhlenbeck process. Section~\ref{sec:synch} is devoted to the analysis of synchronization in the network, whereas Section~\ref{sec:synchstab} addresses the stability of the synchronous state. Finally, Section~\ref{sec:conc} summarizes the main conclusions.

\section{The model}
\label{sec:model}

The Kuramoto model with inertia is used to describe the phase and frequency dynamics of a set of interconnected generators, such as wind turbines within a power network~\cite{taher2019enhancing}. In this framework, each generator is represented as a synchronous machine characterized by a phase angle $\delta$ and an angular velocity $\varphi=d\delta/dt$ defined with respect to a rotating reference frame at the network reference frequency $\omega_R = 2 \pi f$ with $f = 50~\mathrm{Hz}$. Following ref.~\cite{Taher2019}, the dynamics of each generator $i$ are described by the ordinary differential equation given in Eq.~\eqref{eqt1}:
%~\cite{Taher2019}:
\begin{equation}
\begin{cases}
 \dfrac{d{\delta}_i}{dt} &= \varphi_i \\
\dfrac{d{\varphi}_i}{dt} &= -{D_{wind_i}} \varphi_i + \dfrac{P_{wind_i}{(t)}}{H_{wind_i}\omega_R} + \dfrac{K }{H_{wind_i}\omega_R} \sum\limits_{\substack{j=1 \\ j \ne i}}^N A_{ij} \sin(\delta_j - \delta_i)\, .
\end{cases}
\label{eqt1}
\end{equation}
The mechanical power produced by the $i$-th wind turbine is modeled {as shown in Eq.~\eqref{eqt1m}}~\cite{stiebler2008wind}: 
\begin{equation}
    {P_{wind_i}(t)}={\frac{1}{2}\rho S C_p V_i^3(t)}\, ,
    \label{eqt1m}
\end{equation} 
where $\rho$ represents the air density, $S$ the swept area of the wind turbine blades, $C_p = 0.59$ the performance coefficient of the wind turbine~\cite{kavade2025variable}, $H_{wind_i}$ the moment of inertia of generator $i$ and ${D_{wind_i}}$ models the energy losses due to friction and electromechanical damping of the $i$-th generator, with $i = 1,...,N$. $K>0$ is the coupling coefficient between the generators, $A_{ij}$ is the $(i,j)$ the elements of the adjacency matrix describing the topology of the network, namely how the turbine are connected each other and $ \sum\limits_{\substack{j=1 \ j \ne i}}^N A_{ij} \sin(\delta_j - \delta_i)$ represents the interactions between the wind generators. Table~\ref{tab:tableI} summarizes the values of parameters used in this work. The variation of the wind speed over time is described by the term $V_i(t)$; let us observe that we hereby assume the latter to not depend on the turbine index, namely it is constant across the turbine farm. The frequency synchronization is achieved when ${\varphi}_i = 0$ for all $i$ (in a rotating frame with the nominal frequency $50~\mathrm{Hz}$), indicating that all turbines oscillate at the same frequency as the one of the network, without phase deviation. The moment of inertia of a wind turbine $H_{wind}$ constitutes a fundamental dynamic parameter, since it determines the ability of the rotor to resist rapid speed variations imposed by the wind. This inertia mainly originates from the blades whose contribution dominates due to their mass located at a large distance from the axis and from the hub, more compact but massively concentrated near the center. According to ref.~\cite{heier2014grid}, the total inertia referred to the low-speed shaft is expressed as:
\begin{equation}
    H_{rotor} = H_{\text{blades}} + H_{\text{hub}}\, .
\end{equation}
In geared wind turbines, this inertia is transmitted to the generator through the gear ratio $G$, which is equal to the ratio between the rotational speed of the wind generator and that of the rotor (turbine). The equivalent inertia seen by the wind generator is then defined by {Eq.~\eqref{eqtm1h} (see ref.~\cite{gonzalez2023unveiling} for details)}:
\begin{equation}
  H_{\text{wind}} = \frac{H_{\text{rotor}}}{G^{2}},
  \quad with, \quad
  G = \frac{\omega_{\text{generator}}}{\omega_{\text{rotor}}}\, .
  \label{eqtm1h}
\end{equation}
%%%%%%%%%%%%%%%%%%%%%%%%
%\begin{equation}
%  H_{wind} = \frac{H_{\text{rotor}}}{G^{2}}\, ,
%  \label{eqtm1h}
%\end{equation} 
%with 
%\begin{equation}
%  G = \frac{\omega_{\text{generator}}}{\omega_{rotor}}\, .
%\end{equation} 
%%%%%%%%%%%%%%%%%%%%%%%%
This relation shows that the gearbox greatly reduces the apparent inertia of the rotor on the high-speed shaft, which directly influences the electromechanical dynamics of the machine, its stability, and its response to network disturbances. In the case of the studied wind network, the parameters studied are listed in Table~\ref{tab:tableI}.
\begin{table*}[ht]
\centering
\caption{Physical units of the wind turbine model parameters~\cite{Taher2019,Menck2014}.}
\begin{tabular}{lccc}
\hline
\textbf{Parameter} & \textbf{Symbol} & \textbf{values} & \textbf{Unit (SI)}  \\
\hline
Rotor diameter & $d$ & 23.2 & m \\
Moment of inertia & $H_{wind}$   & $40$ & kg·m$^2$ \\
Mechanical damping coefficient & $D_{wind}$  & $0.5$ & rad·s$^{-1}$ \\
Network reference frequency & $\omega_R = 2\pi f$   & $314$ & rad·s$^{-1}$ \\
Power produced by the wind turbine & $P_{wind}$  & - & W  \\
Wind turbine performance coefficient  & $C_p$  & $0.59 $ &  (dimensionless) \\
Average wind speed & $V_{mean}$  & $3$ & m·s$^{-1}$ \\
Air density & $\rho$  & $1.22$ & kg·m$^{-3}$ \\
Area swept by the wind turbine blades & $S = \dfrac{\pi d^2}{4}$  & $422.51$ &   m$^2$ \\
Rotor phase & $\delta$  & - & rad \\
Angular velocity of the rotor phase & $\varphi$  & - & rad·s$^{-1}$ \\
Angular acceleration of the rotor phase  & $\dot{\varphi}$ & - & rad·s$^{-2}$ \\
Network adjacency matrix (unweighted)  & $A_{ij}$  & - &  (dimensionless)\\
Coupling strength between the nodes  & $K$  & $4\times 10^{3}$ & $W$ \\
\hline
\end{tabular}
%\caption{Physical units of the wind turbine model parameters~\cite{Taher2019,Menck2014}. \Th{Le titre d’un tableau se met au-dessus du tableau.}}
\label{tab:tableI}
\end{table*}

\subsection{Modeling wind speed by using the Ornstein–Uhlenbeck process}
\label{ssec:OUproc}
Wind speed exhibits random fluctuations, with a non-negligible temporal correlation due to atmospheric phenomena. To realistically represent this behavior, we adopt an Ornstein–Uhlenbeck (OU) stochastic process~\cite{maller2009ornstein,martinez2023dynamical, jonsdottir2020stochastic,pham2025application}. Unlike a purely static modeling based solely on a probabilistic distribution such as the Weibull distribution~\cite{justus1978methods, takle1978characteristics}, the OU process allows the introduction of a correlated temporal dynamics, which is essential for studying the stability and robustness of wind generators in the face of realistic wind fluctuations, description that cannot achieved with wind speed described by the Weibull distribution~\cite{luna2012optimal}. The Ornstein-Uhlenbeck process~\cite{pham2025application}, widely used to model mean-reverting random phenomena, is defined by the following stochastic {ordinary} differential equation {given in Eq.~\eqref{eq:x1}}:
\begin{equation}
d\xi^{\text{ou}}(t) = -\frac{1}{\tau_{\text{ou}}}\,\xi^{\text{ou}}(t)\,dt
+ \sqrt{\frac{1}{\tau_{\text{ou}}}}\, dW(t),
\label{eq:x1}
\end{equation}
where $\tau_{\text{ou}}$ represents the wind correlation time and $W(t)$ is a standard Brownian motion modeling the random turbulent excitation, characterized by Gaussian increments with zero mean and unitary standard deviation. This stochastic process is characterized by an exponential temporal correlation~\cite{martinez2023dynamical} given by Eq.~\eqref{eq:OU_continu}:
\begin{equation}
\langle \xi^{\text{ou}}(t)\xi^{\text{ou}}(t') \rangle
= \frac{1}{2\tau_{\text{ou}}}
\,e^{-|t-t'|/\tau_{\text{ou}}}\, .
\label{eq:OU_continu}
\end{equation}
The instantaneous wind speed is therefore modeled according to Eq.~\eqref{eq:OU_wind}, following~\cite{martinez2023dynamical,tchuisseu2017effects}:
\begin{equation}
V(t) = V_{mean} \left( 1 + \epsilon_{\text{ou}}\, \xi^{\text{ou}}(t) \right)\, ,
\label{eq:OU_wind}
\end{equation}
where ${V}_{mean}$ denotes the instantaneous mean wind speed, $\epsilon_{\text{ou}}$ denotes the relative intensity of wind speed fluctuations, and $\xi^{\text{ou}}(t)$ the OU process. In this work, the fluctuation intensity is fixed at $\epsilon_{\text{ou}} = 0.15$ and the wind correlation time $\tau_{\text{ou}}=60\, s$~\cite{martinez2023dynamical}, corresponding to moderate variations but reflecting the realistic aspect of wind speed around its mean value. 

For the numerical solution, the continuous Eq.~\eqref{eq:x1} is solved analytically over a discrete time interval $[t_k, t_{k+1}]$ of duration $\Delta t$. By integrating the Ornstein–Uhlenbeck equation {as in ref.~\cite{gillespie1996exact}} over this interval, the following discretized solution is obtained:
\begin{equation}
\xi_{k+1}^{\text{ou}} =
e^{-\Delta t/\tau_{\text{ou}}}\,\xi_{k}^{\text{ou}}
+ \sqrt{\frac{1-e^{-2\Delta t/\tau_{\text{ou}}}}{2\tau_{\text{ou}}}}
\,\eta_{k}\,,
\label{eq:OU_discret}
\end{equation}
where $\eta_{k}$ is a Gaussian random variable with zero mean and unit variance. This formulation corresponds to the exact analytical solution of the OU process evaluated at discrete time points and allows the temporal correlation structure of the wind to be clearly preserved.
This modeling thus enables the generation of continuous, realistic, and temporally correlated wind trajectories, which are essential for analyzing the instantaneous dynamics of the synchronous point and assessing the impact of wind fluctuations on wind power generation and the overall stability of the electrical grid.

\begin{figure}[htp!]
\centering
\begin{tabular}{cc}
     \includegraphics[width=0.5\linewidth]{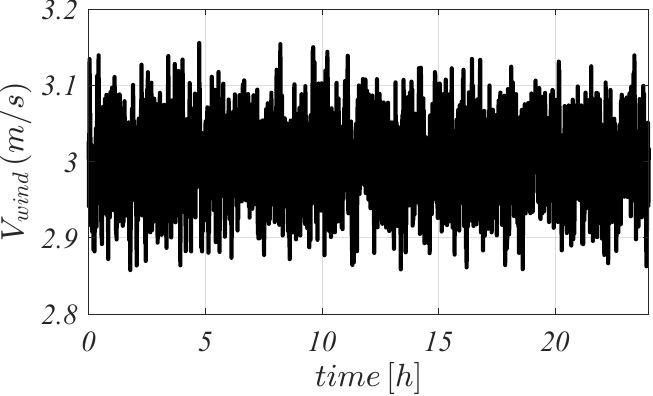}&  \includegraphics[width=0.42\linewidth]{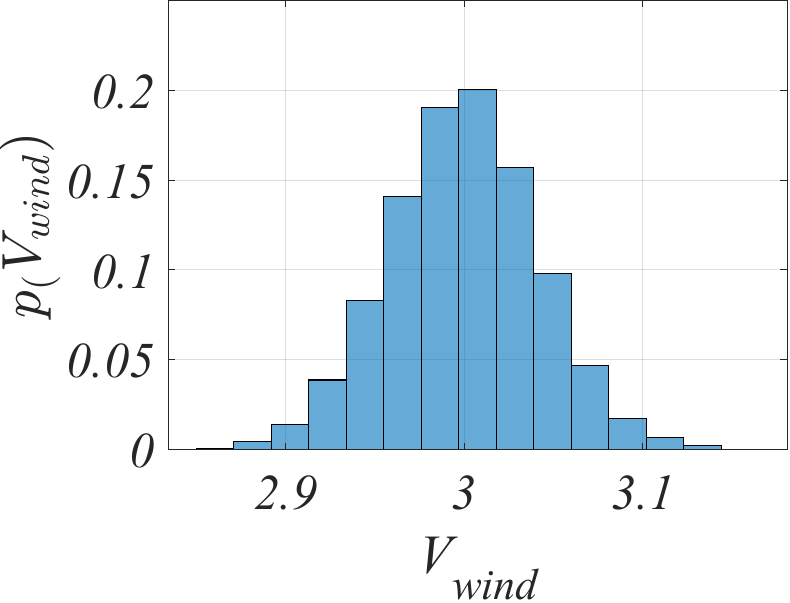}\\
     (a) & (b)
\end{tabular}
\caption[]{%
\protect\justifying
\textbf{Time evolution of wind speed}. In panel~\Th{(a)} we report the temporal evolution of the wind speed obtained from a numerical integration of the OU process. Panel~\Th{(b)} shows the distribution of the wind speeds. The average speed is $V_{\text{mean}} = 3~\mathrm{m/s}$. } 
\label{fig:figure1}
\end{figure}

To apply this model to a concrete context, it is necessary to rely on local wind measurements. {In the specific context of this study, we used an average wind speed $V_{\text{mean}}$ corresponding to measurements from the city of Namur, Belgium, of approximately $3\ \mathrm{m/s}$, based on long-term climatological data reported by the Royal Meteorological Institute of Belgium~\cite{irm_climate}. Unless otherwise specified, this value is adopted throughout the work.} Panel~(a) of Figure~\ref{fig:figure1} shows a generic time series of wind speed obtained by simulating the Ornstein--Uhlenbeck process. The curve displays random and irregular fluctuations, characteristic of weakly turbulent wind conditions~\cite{ren2018analysis}. 
%%%%%%%%%%%%%%%%%%%%%%%%%%%%%%%%
%in the case of Belgium, and more precisely in the city of Namur, the average wind speed $V_{\text{mean}}$ was measured around $3\ \mathrm{m/s}$; this is the value we will use throughout this work, unless differently specified. In panel (a) of Figure~\ref{fig:figure1} we display a generic time series representing the wind speed obtained by simulating the OU process; the curve shows random and irregular variations, characteristic of a wind of weak turbulent nature~\cite{ren2018analysis}. 
%%%%%%%%%%%%%%%%%%%%%%%%%%%%%%%%%%%%%%%
These fluctuations may result from rapid meteorological changes or from local conditions (relief, obstacles, etc.). In panel~(b) we show the distribution of the wind speeds for the values of panel~(a).

\section{Synchronization for a network of wind turbines}
\label{sec:synch}
Let us consider a network of $50$ coupled wind generators, all sharing identical parameters, except for inertia. We assume the wind farm to be small enough for all generators to receive the same wind speed at each instant. The network dynamics is modeled according to {Eq.}~\eqref{eqt1}. To analyze the synchronization of the network, we examine the evolution of the key variables of the system, {namely} the angles $\delta_i$, and the angular velocities $\varphi_i$, {together with the} evolution of the order parameter $R$ {expressed by Eq.~\eqref{eqr}} 
\begin{equation}
R(t) = \Big|\frac{1}{N} \sum_{i=1}^{N} e^{j \delta_i(t)}\Big|\, ,
\label{eqr}
\end{equation}
and the standard deviation of the frequencies $\sigma_\varphi$
\begin{equation}
\sigma_\varphi(t) = \sqrt{\frac{1}{N} \sum_{i=1}^{N} \left( \varphi_i(t) - \bar{\varphi}(t) \right)^2}\, ,
\end{equation}
 with $\bar{\varphi}(t) = \frac{1}{N} \sum_{i=1}^{N} \varphi_i(t)$. The former quantifies the degree of coherence of the angles, for $R(t) \approx 1$ the phases are almost identical (maximum coherence or synchronization),
$R(t) \approx 0$, the phases are dispersed (desynchronization). The latter makes it possible to measure the degree of dispersion around the average frequency $\bar{\varphi}(t)$. We will be particularly interested in studying the influence of the average wind speed, the coupling strength, as well as inertia and damping on synchronization. {To illustrate these effects and provide insight into the dynamical behavior of the system, we first examine the temporal evolution of the generator states under representative operating conditions.}

\begin{figure}[htp!]
    \centering
        \begin{tabular}{cc}
            \includegraphics[scale=0.37]{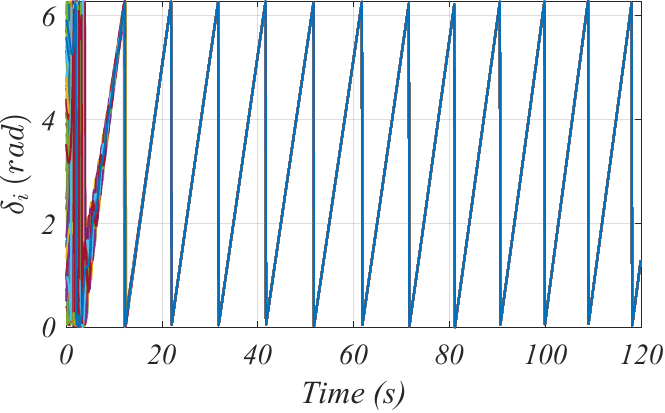} &
            \includegraphics[scale=0.37]{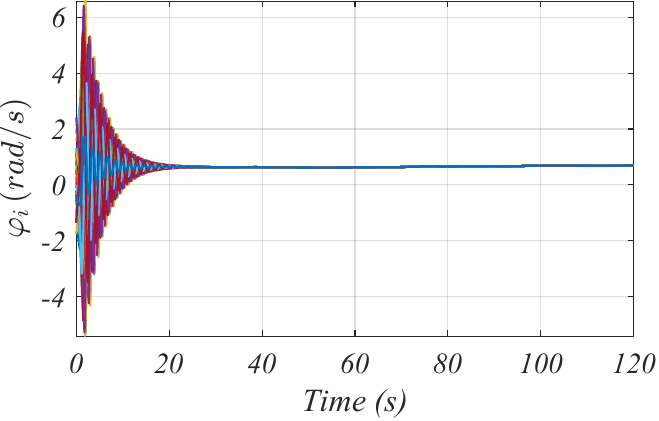}\\
            (a) & (b) \\
            \includegraphics[scale=0.37]{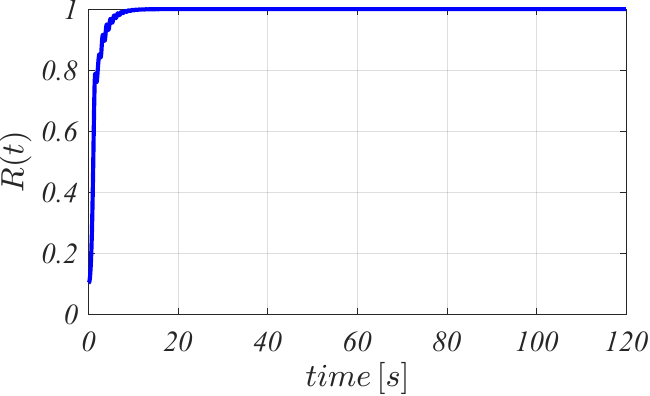} &
            \includegraphics[scale=0.37]{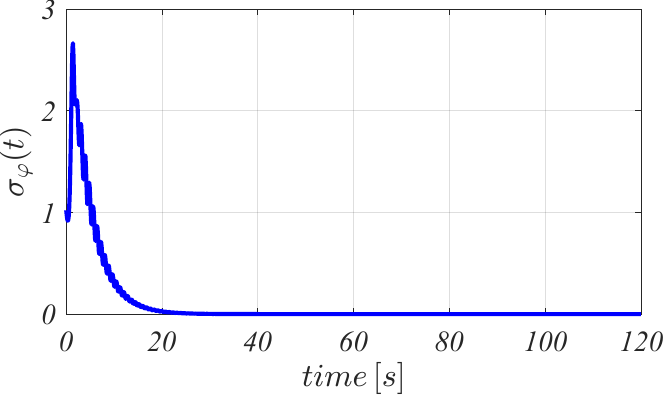}\\
            (c) & (d) \\
        \end{tabular}
        \caption[]{%
        \protect\justifying
\textbf{Synchronization of coupled wind turbines.}
Panel~(a) reports the temporal evolution of the wind turbine phases, while panel~(b) shows the evolution of their rotational speeds. Panel~(c) illustrates the time evolution of the order parameter, $R(t)$, and panel~(d) depicts the standard deviation of the frequencies, $\sigma_{\varphi}(t)$. The parameters used are~\cite{Taher2019,Menck2014}: $N=50$; the turbine inertia $H_{\mathrm{wind}_i}$ is uniformly distributed in the interval $[40,40.05]~\mathrm{kg\,m^2}$; the coupling strength is $K=4\times10^3~\mathrm{W}$; the damping coefficient is $D_{\mathrm{wind}_i}=0.5~\mathrm{rad/s}$ for all $i=1,\dots,N$; the reference angular frequency is $\omega_R=2\pi f~\mathrm{rad/s}$ with $f=50~\mathrm{Hz}$; and the mean wind speed is $V_{\mathrm{mean}}=3~\mathrm{m/s}$.
%%%%%%%
}
\label{fig:figure2}
\end{figure}

In Figure~\ref{fig:figure2} panel~(a) we show the evolution of the rotor angles $\delta_i$ of the different generators in the network as a function of time.
It can be observed that, despite initial differences, the angles gradually adopt a common behavior, namely the synchronous state, where all generators oscillate at the same frequency and maintain a constant phase shift relative to one another. In Figure~\ref{fig:figure2} panel~(b)
we present the evolution of the angular velocities $\varphi_i$ (or frequencies) of the generators. It shows a progressive convergence of $\varphi_i$ to a constant value. We can thus conclude that the system of wind turbine converges to complete synchronization. The same conclusion can be drawn by looking at the time evolution of the order parameter $R(t)$ (panel~(c)) and the dispersion of frequencies among the generators
through the standard deviation (panel~(d)). It can be observed that the deviation increases rapidly
to reach approximately $2.5~\mathrm{rad/s}$, then gradually decreases and stabilizes
at zero after about $20$ seconds, indicating frequency coherence among the generators.

\begin{figure}[htp!]
    \centering 
    \begin{tabular}{cc}
        \includegraphics[width=0.5\linewidth]{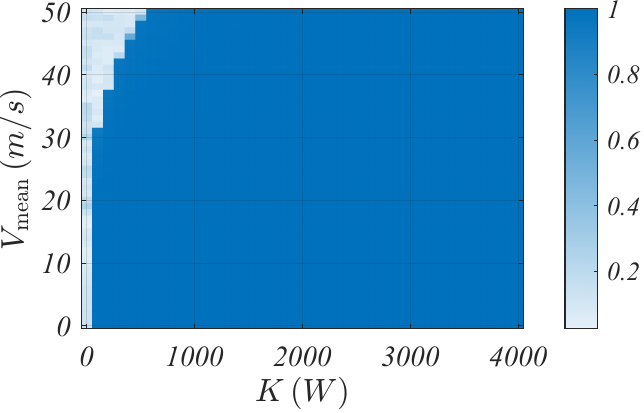} &  
        \includegraphics[width=0.5\linewidth]{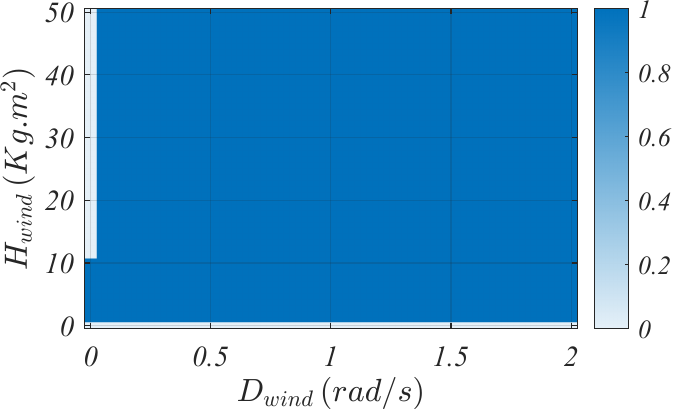}\\
         (a) & (b)
    \end{tabular}
    \caption[]{%
    \protect\justifying \textbf{Synchronization analysis of a network of interconnected wind generators}. Panel (a) presents the order parameter as a function of the average wind speed and the coupling strength $K$. The wind-generator inertia $H_{\mathrm{wind}}$ is assumed to follow a uniform distribution in the range $[40,40.05]~\mathrm{kg.m^2}$, with damping coefficient $D_{\mathrm{wind}} = 0.5~\mathrm{rad/s}$. Panel (b) illustrates the dependence of the order parameter on inertia and damping for a fixed coupling strength of $K = 4 \times 10^{3}~\mathrm{W}$. The remaining parameters are set to $N = 50$ and $V_{\mathrm{mean}} = 3~\mathrm{m/s}$. }
    \label{fig:figure3}
\end{figure}

%%%%%%%%%%%%%%%%%%%%%%%%%%
%%%%%%%%%%%%%%%%%%%%%
{Subsequently, we analyze how the average wind speed $V_{mean}$ and coupling strength $K$, together with the damping coefficient $D_{wind}$ and inertia $H_{wind}$, influence the synchronization dynamics of the network. This is assessed through the behavior of the order parameter across the parameter spaces ($K$, $V_{mean}$) and ($D_{wind}$, $H_{wind}$).}
%%%%%%%%%%%%%%%%%%%%%%%%%%%%%%%%
%%%%%%%%%%%%%%%%%%%%%%%%%%%%
%Subsequently, we focus on the impact of the average wind speed, the coupling strength between the generators, the inertia, and the damping coefficient on the synchronization dynamics within the network. 
%%%%%%%%%%%%%%%%%%%%%%%%%%%
%%%%%%%%%%%%%%%%%%%%%%%%%%%
Figure~\ref{fig:figure3} (a) illustrates the {combined} effect of the coupling strength $K$  and the average wind speed $V_{mean}$ on the network synchronization. Synchronization is defined by using the order parameter, namely once $R\rightarrow 1$ (blue region); small values of $R$ are reported in white. Let us observe that synchronization is almost always achieved except for very small value of $K$ or large values of ${V}_{mean}$. We can thus conclude that strengthening the coupling between the wind generators promotes the stability of the network, the stronger the wind the larger the coupling to achieve synchronization. { This suggests a competition between coupling-induced coherence and wind-induced fluctuations, which tend to disrupt collective dynamics. As a consequence, stronger coupling is required to counteract the desynchronizing effect of high wind conditions and sustain network coherence.}
Figure~\ref{fig:figure3} (b) displays the effect of inertia $H_{wind}$ and the damping coefficient $D_{wind}$, on network synchronization. We can observe that synchronization (blue region) always emerges except for very small $D_{wind}$ or very low values for $H_{wind}$. {Therefore, these results suggest that, within the considered range, the synchronization dynamics are weakly sensitive to variations in wind inertia and damping. In contrast to the influence of coupling strength and average wind speed illustrated in panel (a), these parameters do not play a dominant role in the loss of synchrony, highlighting the overall robustness of the network with respect to such parameters dynamical effects.}
%%%%%%%%%%%%%%%%%%%%

\section{Stability of the synchronous state}
\label{sec:synchstab}
This section focuses on the study of the stability of the synchronous state for a single wind generator coupled to a stable grid, facing to significant disturbances. Indeed in a real network, disturbances can be substantial, such as: the loss of a generator, a wind gust, a sudden increase in load, or a network fault.
In such cases, linear stability is no longer sufficient, and it is necessary to measure the global stability of the network. The dynamic of a {single} wind generator {characterized by its} phase $\delta$ and frequency $\varphi$ is given by the following Eq.\eqref{eqt2}:
\begin{equation}
\begin{cases}
\dfrac{d{\delta}}{dt} = \varphi \\
\dfrac{d{\varphi}}{dt} = -{D_{wind}} \varphi + \dfrac{P_{wind}(t)}{H_{wind}\omega_R} - \dfrac{K}{H_{wind}\omega_R} \sin(\delta)\, ,
\end{cases}
\label{eqt2}
\end{equation}
where we have assumed the phase of the grid to vanish, i.e., $\delta_{\text{grid}} = 0$. 
%%%%%%%%
To simplify the notations, let us rewrite the previous system as in Eq.\eqref{eq:systeme1}:
\begin{equation}
\begin{cases}
\dfrac{d{\delta}}{dt} = \varphi\\
\dfrac{d{\varphi}}{dt} = -\alpha \varphi + P(t) - k \sin(\delta - \delta_{grid})\, ,
\end{cases}
\label{eq:systeme1}
\end{equation}
where $\alpha={D_{wind}}$, $P(t)=\dfrac{P_{wind}(t)}{H_{wind}\omega_R}$, and $k=\dfrac{K }{H_{wind}\omega_R}$.

Let us now consider the entire system (the wind generator and the grid) to lie in a synchronous state~\cite{menck2013basin}, then we perturb the initial conditions of the wind generator and we determine if the latter converges back to the synchronous state. We hence define the basin of stability, $B$, of the synchronous state as the set of initial conditions whose orbit eventually converges to the synchronous state. 
%%%%%%%%%%%%%
We are interested in studying the size of the basin, $S(B)$, as a function of the main model parameters, i.e., $K$, $D_{wind}$ and $H_{wind}$. We hereby estimate the basin stability by using a Monte Carlo-type numerical procedure~\cite{menck2013basin}, which determines the fraction of initial conditions that return to the synchronous state. This analysis will be divided into two parts; first we will consider the variation in wind speed to be negligible so to consider the wind constant. In a second part, we will model the wind as a stochastic OU process as described above.

\subsection{Stability of wind generator with constant wind power}
\label{ssec:constwind}
Because we are assuming wind to be constant, also the power created by the turbine will be constant and thus the synchronous solution of Eq.~\eqref{eq:systeme1} can be straightforwardly determined to return
\begin{equation}
  \delta_{\text{syn}} = \arcsin\left(\frac{P}{k}\right)\quad\text{ and }\quad\varphi_{\text{syn}} = 0\, .
  \label{eqt4}
\end{equation}
Let us observe that this solution exists if and only if $|P| < k$. The idea is to perturb the wind generator from the synchronous state $(\delta_{syn}, 0)$ by adding a perturbation $(\delta_0, \varphi_0)$, compute the orbit with the new initial condition, $(\delta, \varphi)=(\delta_{syn}+\delta_0, \varphi_0)$, and check if it converges back to $(\delta_{syn}, 0)$. In this case {$(\delta, \varphi)$} belongs to the stability basin of the synchronous state.

\begin{figure}[htp!]
    \centering
    \begin{tabular}{cc}
         \includegraphics[height=2.95cm]{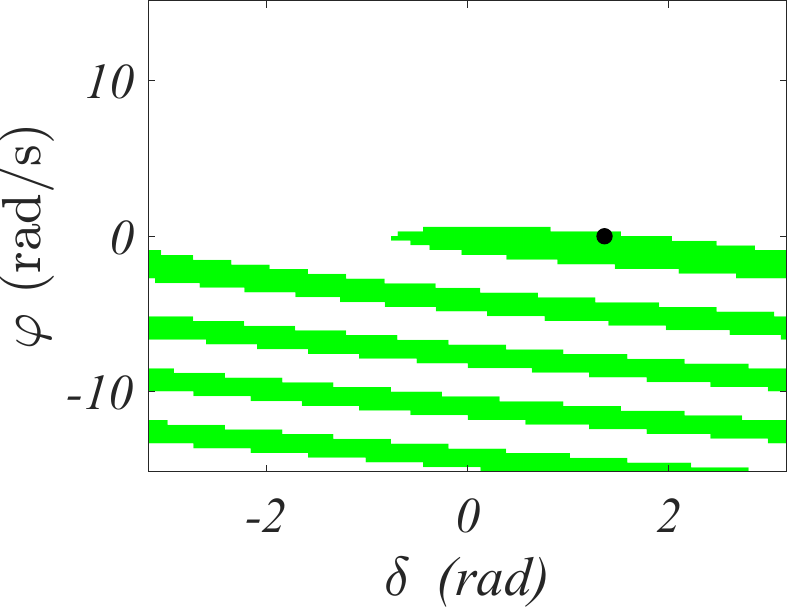}&  
         \includegraphics[height=2.95cm]{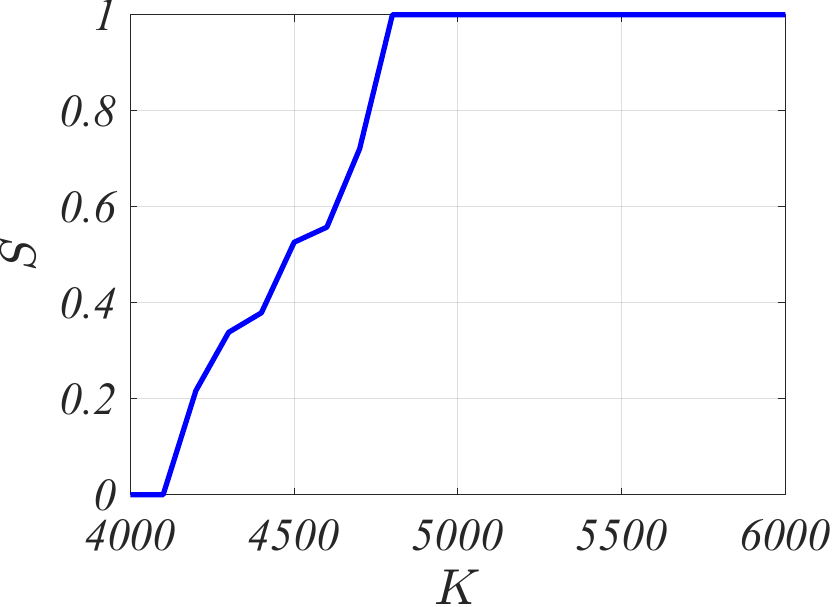}
    \end{tabular}
   \caption[]{%
    \protect\justifying 
    \textbf{Stability basin of the synchronous solution as a function of the coupling strength $K$.}
By using the method presented in the text, we show the set of initial conditions whose trajectories converge to the synchronous solution $(\delta_{\mathrm{syn}},0)$. {In panel~(a), the green region denotes the basin of attraction of the synchronous state, the white region corresponds to initial conditions leading to desynchronization, and the black dot indicates the stable synchronous equilibrium point, shown here for $K=4200~\mathrm{W}$.} Panel~(b) represents the size of the synchronization basin, $S(B)$, as a function of $K$. The other parameters are $D_{\mathrm{wind}}=0.5~\mathrm{rad/s}$, $H_{\mathrm{wind}}=40~\mathrm{kg.m^2}$, and $V_{\mathrm{mean}}=3~\mathrm{m/s}$.}
    \label{fig:figure4}
\end{figure}

\begin{figure}[htp!]
    \centering
    \begin{tabular}{cc}
         \includegraphics[height=2.95cm]{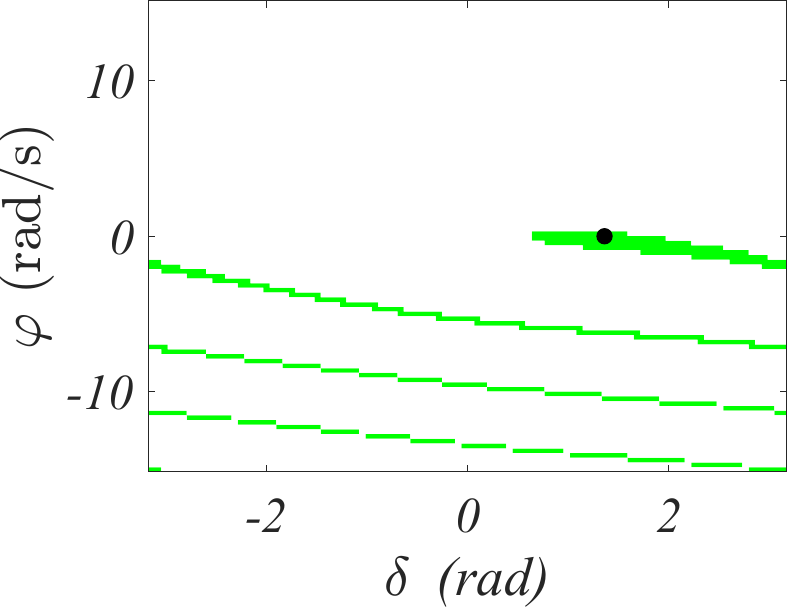}&
         \includegraphics[height=2.95cm]{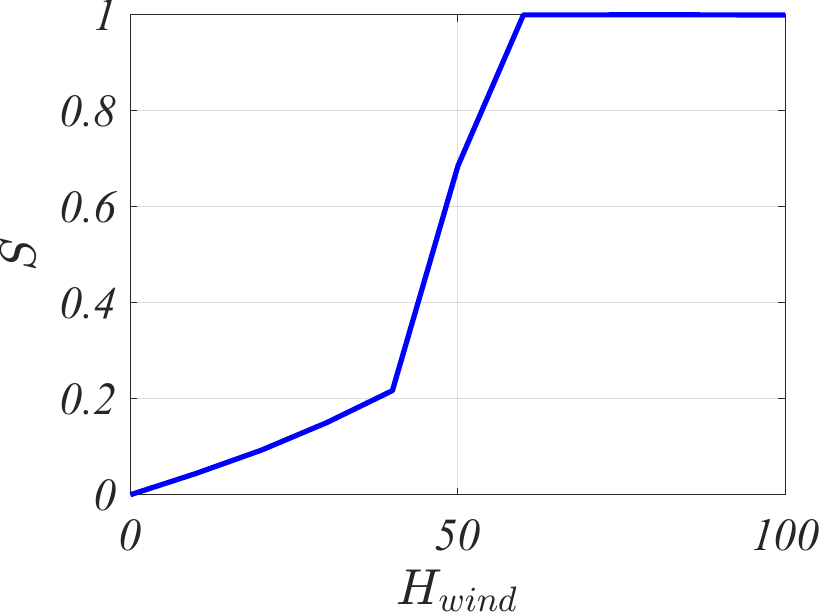}
    \end{tabular}
    \caption[]{%
    \protect\justifying 
    \textbf{Stability basin of the synchronous solution as a function of the inertia parameter $H_{wind}$}.
    Panel~(a) represents {the basin of attraction of the synchronous solution for $H_{\mathrm{wind}}=10~\mathrm{kg.m^2}$. The colors and their meanings are the same as those defined in Figure~\ref{fig:figure4}.} Panel~(b) represents the size of the synchronization basin, $S(B)$, as a function of $H_{\mathrm{wind}}$. The other parameters are $D_{\mathrm{wind}}=0.5~\mathrm{rad/s}$, $K=4200~\mathrm{W}$, and $V_{\mathrm{mean}}=3~\mathrm{m/s}$.}
    \label{fig:figure5}
\end{figure}

{To further characterize the stability properties of the system, we compute and illustrate} in Figure~\ref{fig:figure4} the stability basin of the synchronous solution of a wind turbine generator. Panel~(a) shows the stability basin in the phase space $(\delta,\varphi)$ for a coupling strength $K = 4.2 \times 10^3\ \mathrm{W}$. {In the figure, the green regions correspond to initial conditions that converge toward the synchronized operating state, while the white regions represent the initial conditions that fail to recover synchronization. The black dot denotes the stable synchronous equilibrium point. The basin exhibits a banded structure, reflecting the $multistable$ nature of the system and the strong dependence of the asymptotic dynamics on the initial conditions. This result shows that, synchronization is only achieved for specific intervals of initial phase and frequency, indicating that the synchronous state is only locally stable for the specific value of the coupling strength $K = 4.2 \times 10^3\ \mathrm{W}$. Panel~(b) shows the evolution of the normalized basin size $S$ as a function of the coupling strength $K$ computed using the following Eq.~\eqref{eqbas}:
\begin{equation}
    S(B)=\frac{1}{N_{\mathrm{tot}}}
\sum_{j=1}^{N_{\mathrm{tot}}}
\chi_{B}\!\left(\mathbf{x}^{(j)}\right),
    \label{eqbas}
\end{equation}
where, $N_{\mathrm{tot}}$ is the total number of initial conditions considered in the phase space and the indicator function $\chi_{B}\!\left(\mathbf{x}\right)$ assumes the value one if the trajectory starting from $\mathbf{x}=(\delta,\varphi)$ (initial conditions)
converges to the synchronous state, and zero otherwise.
The results show that for low values of $K$, the basin size is nearly zero, meaning that synchronization cannot be sustained for the considered initial conditions. As the coupling strength increases, $K \geq 4.1\times 10^3~\mathrm{W}$, the size of the synchronization basin starts growing progressively, revealing that a larger portion of the phase space converges toward the synchronous state. Around $K \approx 4.8 \times 10^3~\mathrm{W}$, the basin size reaches the full size, $S=1$, indicating that all tested initial conditions evolve toward synchronization. This transition demonstrates that stronger coupling substantially enhances the robustness and global stability of the synchronized regime in the wind-turbine network.
}
{We then investigate the role of inertia in shaping the stability properties of the wind turbine generator. Figure~\ref{fig:figure5} shows the effect of the wind-turbine inertia $H_{\mathrm{wind}}$ on the basin of attraction of the synchronous state. Panel~(a) displays the synchronization basin in the phase space $(\delta,\varphi)$ for a representative value of the inertia fixed at $H_{\mathrm{wind}} =10~\mathrm{kg.m^2}$. For consistency with the previous analysis, the same colors and their associated meanings as defined in Figure~\ref{fig:figure4} are used here and remain unchanged for the rest of the results. The resulting basin structure shows that synchronization is reached only from a limited subset of initial conditions, underlining the strong sensitivity of the system dynamics to phase and frequency disturbances. 
%%%%%%%%
In panel~(b), the evolution of $S$ shows the strong effect of the inertia on the stability of the synchronous state, and we can clearly observe that for small inertia values, the basin size remains close to zero ($S \approx 0$), indicating that the synchronous state is weakly stable and highly sensitive to disturbances. As the inertia increases, $S$ progressively expands, showing that a larger fraction of the phase space converges toward synchronization. A fast transition is observed around $H_{\mathrm{wind}}\approx 50~\mathrm{kg.m^2}$, beyond which $S$ reaches unity. This implicitly means that the system becomes more robust, as a higher inertia acts as a dynamic filter that dampens rapid speed fluctuations and limits angular deviations. These results demonstrate that increasing inertia significantly enhances the robustness and stability of the synchronized regime by improving the ability of the system to absorb dynamical fluctuations and resist desynchronizing perturbations.}
\begin{figure}[htp!]
    \centering
    \begin{subfigure}{0.22\textwidth}
        \includegraphics[width=\linewidth]{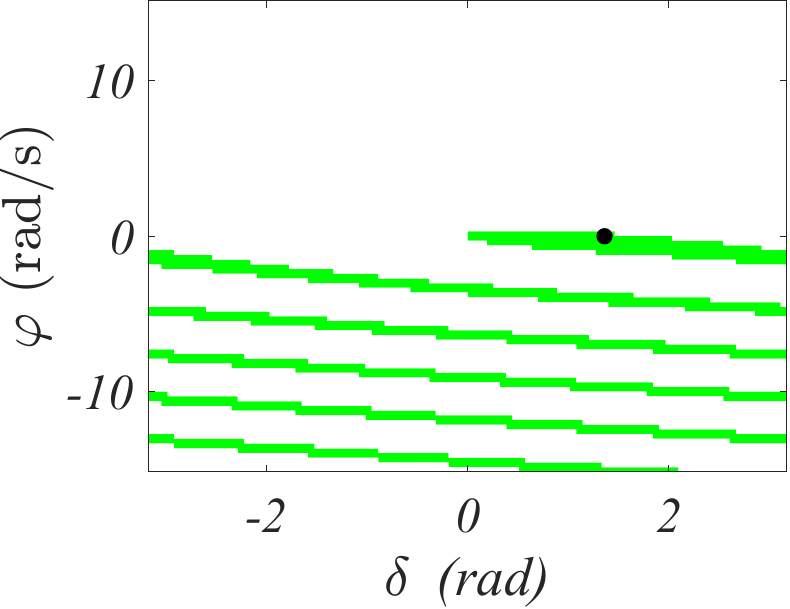}
        \subcaption{}
    \end{subfigure}
    \begin{subfigure}{0.22\textwidth}
        \includegraphics[width=\linewidth]{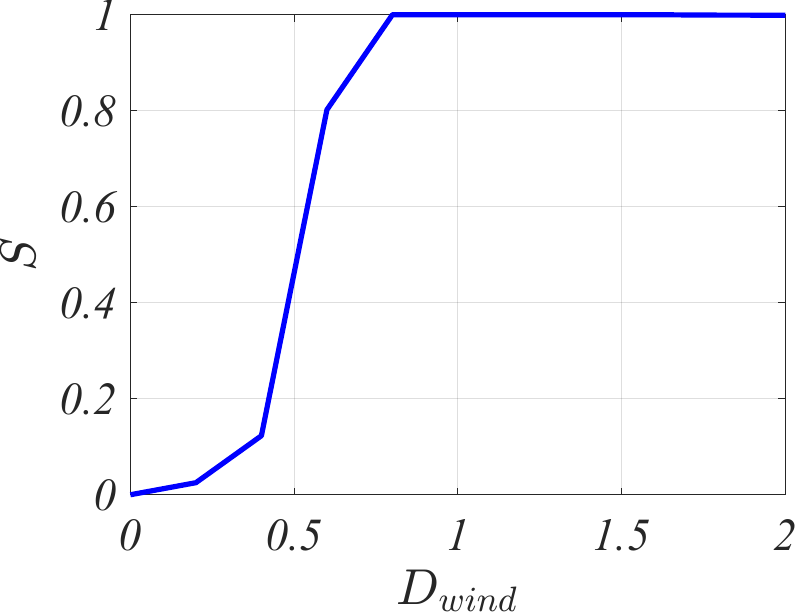}
        \subcaption{}
    \end{subfigure}
    \caption[]{%
    \protect\justifying 
    \textbf{Stability basin of the synchronous solution as a function of the damping parameter $D_{wind}$}.
    {Panel~(a) shows the basin of attraction of the synchronous solution for $D_{\mathrm{wind}}=0.4~\mathrm{rad/s}$. The colors and their meanings are the same as those defined in Figure~\ref{fig:figure4}}. Panel~(b) represents the size of the stability basin, $S(B)$, as a function of $D_{\mathrm{wind}}$. The other parameters are $H_{\mathrm{wind}}=40~\mathrm{kg.m^2}$, $V_{\mathrm{mean}}=3~\mathrm{m/s}$, and $K=4200~\mathrm{W}$.}
    \label{fig:figure6}
\end{figure}
Figure~\ref{fig:figure6} shows the influence of the damping coefficient $D_{wind}$ on the stability of the wind turbine generator 
{through the basin of attraction of the synchronous state. In Panel~(a), we illustrate the basin of attraction of the synchronous solution in the phase space $(\delta,\varphi)$ for a value of the damping coefficient fixed at $D_{\mathrm{wind}}=0.4~\mathrm{rad/s}$. The black dot on this basin represents the stable synchronous equilibrium point. 
}

In the weakly damped regime, the stability region (green domain) remains relatively small and fragmented, indicating a strong sensitivity of the system to perturbations and initial conditions. The generator reacts quickly to disturbances, but the resulting phase and frequency oscillations persist for a long time before the system eventually recovers synchronization. {Since the excess mechanical energy introduced by perturbations is dissipated slowly, the dynamics may exhibit prolonged oscillatory behavior and, for sufficiently large perturbations, evolve toward desynchronization or complex transient dynamics. Panel~(b) shows the evolution of the normalized basin size $S(B)$ as a function of the damping coefficient $D_{\mathrm{wind}}$. As the damping increases, the size of the basin of attraction expands significantly, and this demonstrates that for the considered initial conditions, a larger fraction of the phase space converges toward the synchronous operating state. Around $D_{\mathrm{wind}}\gtrsim 0.75~\mathrm{rad/s}$, the basin size reaches $S(B)=1$, meaning that $100\%$ of the tested initial conditions converge toward synchronization. In real system, increasing the damping coefficient enhances the dissipation of the rotor kinetic energy, thereby reducing the amplitude and duration of oscillations after perturbations. And this induces the reduction of the transient dynamics, which become shorter and more stable, leading to a substantial improvement in the robustness and resilience of the synchronized regime in the interconnected wind-turbine network.
}

\subsection{Stability of the wind generator with variable wind-power} \label{ssec:vartwind}
{We now extend the previous study performed first for a constant mean wind speed and then for a constant wind power, to a more realistic situation in which the wind turbine is subjected to a time-dependent wind power, denoted by $P_{\mathrm{wind}}(t)$. This power depends on the temporal variations of the wind speed noted $V(t)$. As in the previous study, we assume the wind turbine to be coupled to a stable electrical grid, whose reference phase is fixed at $\delta_{\mathrm{grid}}=0$. Under these assumptions, the time evolution of the wind turbine remains governed by Eq.~\eqref{eq:systeme1}.
} The assumption of variable wind power prevents us from obtaining an explicit analytical solution of the system. Consequently, Eq.~\eqref{eq:systeme1} is solved numerically over the time interval $[0,t_{\mathrm{fin}}]$. To evaluate the dynamic coherence of the system under stochastic wind fluctuation, the model is integrated $M$ times, and the mean trajectories of the angle and frequency are then defined by Eq.~\eqref{eqm}:

\begin{equation}
\bar{\delta}(t) = \frac{1}{M}\sum_{j=1}^{M} \delta_j(t)\, ,
\qquad
\bar{\varphi}(t) = \frac{1}{M}\sum_{j=1}^{M} \varphi_j(t)\, .
\label{eqm}
\end{equation}
From the latter, we can compute the variability of each solution with respect to the average phase, namely
\begin{equation}
\varepsilon_{\delta_j} =
\sqrt{
\frac{1}{t_{\mathrm{fin}}}
\int_{0}^{t_{\mathrm{fin}}}
\left( \delta_j(t) - \bar{\delta}(t) \right)^2 dt
}\, ,
\end{equation}
and with respect to the average frequency
\begin{equation}
\varepsilon_{\varphi_j} =
\sqrt{
\frac{1}{t_{\mathrm{fin}}}
\int_{0}^{t_{\mathrm{fin}}}
\left( \varphi_j(t) - \bar{\varphi}(t) \right)^2 dt
}\, .
\end{equation}
We eventually define the total variability, {which} measures of the overall level of dispersion of the trajectories as:
\begin{equation}
\varepsilon_{\mathrm{tot}}^{(1)} =\frac{1}{M}\sum_{j=1}^M\left(
\varepsilon_{\delta_j} + \varepsilon_{\varphi_j} \right)\, .
\end{equation}

Let us analyze the robustness of the synchronous state with respect to perturbations of initial conditions. To this end, we first fix a reference orbit denoted by $\delta_{\mathrm{ref}}(t)$ and $\varphi_{\mathrm{ref}}(t)$, obtained from a given set of initial conditions and defined on the time interval $[0,t_{\mathrm{fin}}]$. We now  perturb the initial conditions by adding $(\delta_0,\varphi_0)$ and we compute the corresponding new solution, denoted by $\delta_{\mathrm{pert}}(t)$ and $\varphi_{\mathrm{pert}}(t)$, over the same time interval.
We measure the difference among the two solutions by defining {the deviation associated with the phase variable and quantified by the root-mean-square error Eq.~\eqref{eqp}}
\begin{equation}
\varepsilon_{\delta}^{\mathrm{pert}} =
\sqrt{
\frac{1}{{\mathrm{\theta}}}
\int_{t_{\mathrm{fin}-{\theta}}}^{t_{\mathrm{fin}}}
\left( \delta_{\mathrm{pert}}(t) - \delta_{\mathrm{ref}}(t) \right)^2 dt}\, ,
\label{eqp}
\end{equation}
{This relation measures the average separation between the perturbed and reference phase trajectories after the transient regime. Similarly, the deviation associated with the frequency variable is defined as:}
%and similarly for the frequency
\begin{equation}
\varepsilon_{\varphi}^{\mathrm{pert}} =
\sqrt{
\frac{1}{{\mathrm{\theta}}}
\int_{t_{\mathrm{fin}-{\theta}}}^{t_{\mathrm{fin}}}
\left( \varphi_{\mathrm{pert}}(t) - \varphi_{\mathrm{ref}}(t) \right)^2 dt}\, ,
\end{equation}
where $\theta$ determines the transient time after which we verify if the system achieved or not synchronization; in the following we will use $\theta=50 \,s$.
The total difference among the reference orbit and the perturbed one is given by Eq.~\eqref{eqpf}:
\begin{equation}
\varepsilon_{\mathrm{tot}}^{(2)} =
\varepsilon_{\delta}^{\mathrm{pert}} + \varepsilon_{\varphi}^{\mathrm{pert}}\, .
\label{eqpf}
\end{equation}

We are now able to measure the impact of initial conditions on system synchronization. If $\varepsilon_{\mathrm{tot}}^{(2)}$ remains small compared to $\varepsilon_{\mathrm{tot}}^{(1)}$, then the system can be considered robust and dynamically stable. Conversely, a significant increase in the total error after perturbation indicates strong dynamic sensitivity and may signal a loss of synchronization or a transient instability. 

{Using these metrics, we can therefore investigate the basin of synchronization in this more realistic operating condition of wind fluctuations. They provide a quantitative criterion to distinguish between stable and unstable responses under wind fluctuations. More precisely, as in the previous section, in the following results of this work, we consider} trajectories whose deviations remain within the prescribed error band and whose frequency deviation converges to zero as stable and synchronous; these trajectories are represented by the green regions. This behavior indicates the capability of the system to restore the nominal frequency after a disturbance. Conversely, trajectories that remain bounded within the error band but whose frequency deviation does not converge to zero are classified as stable but unsynchronized, and are represented by the white regions. In this case, although the dynamical variables do not diverge, the system does not recover the nominal operating frequency of $50~\mathrm{Hz}$, which indicates a loss of synchronism from the viewpoint of grid operation. This distinction allows for a precise identification of the regions of the basin that lead to a state that is both stable and synchronous, corresponding to the normal operating regime of the electrical network.
\begin{figure}[htp!]
    \centering
\begin{tabular}{cc}
     \includegraphics[width=0.235\textwidth]{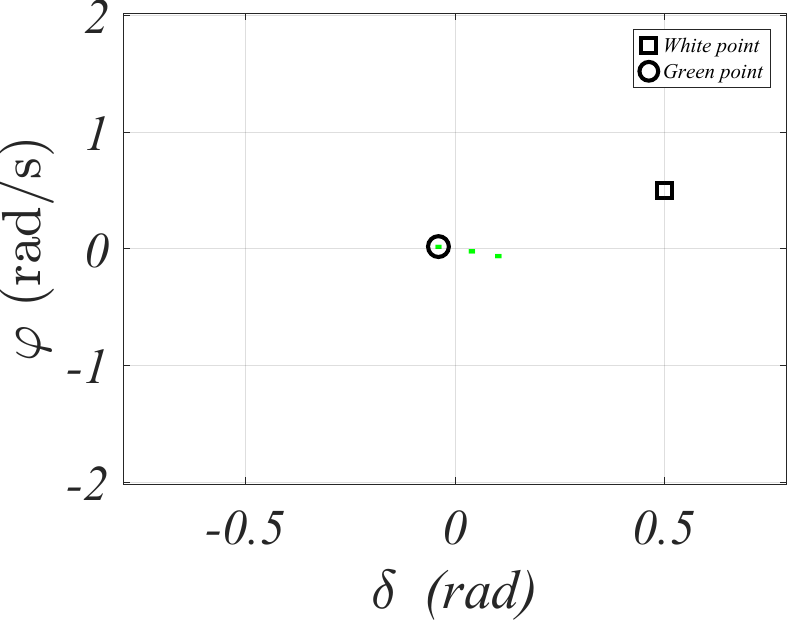}& 
     \includegraphics[width=0.235\textwidth]{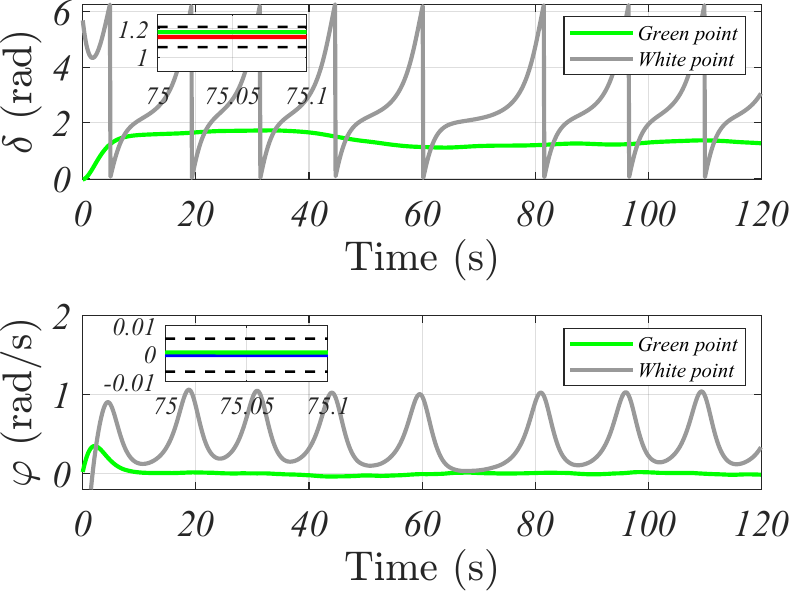}\\
     (a) & (b) \\
     \includegraphics[width=0.235\textwidth]{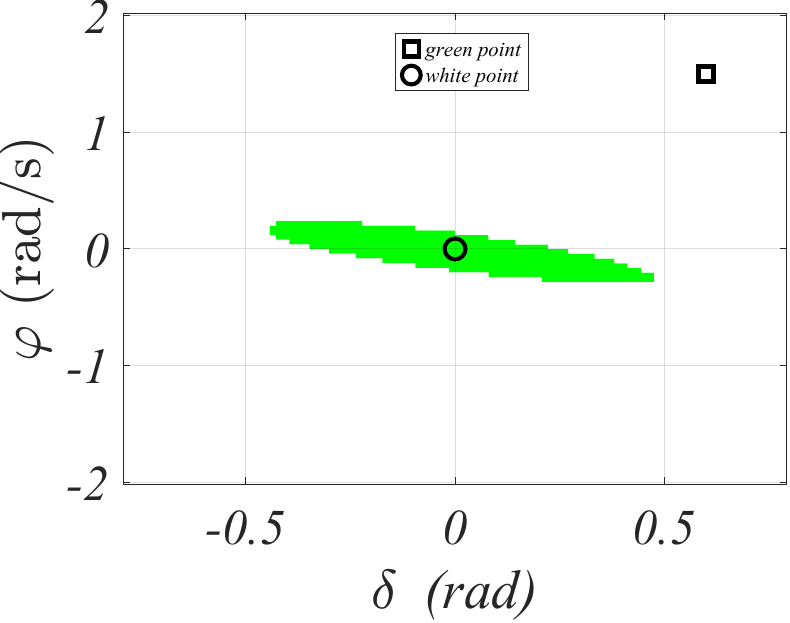}& 
     \includegraphics[width=0.235\textwidth]{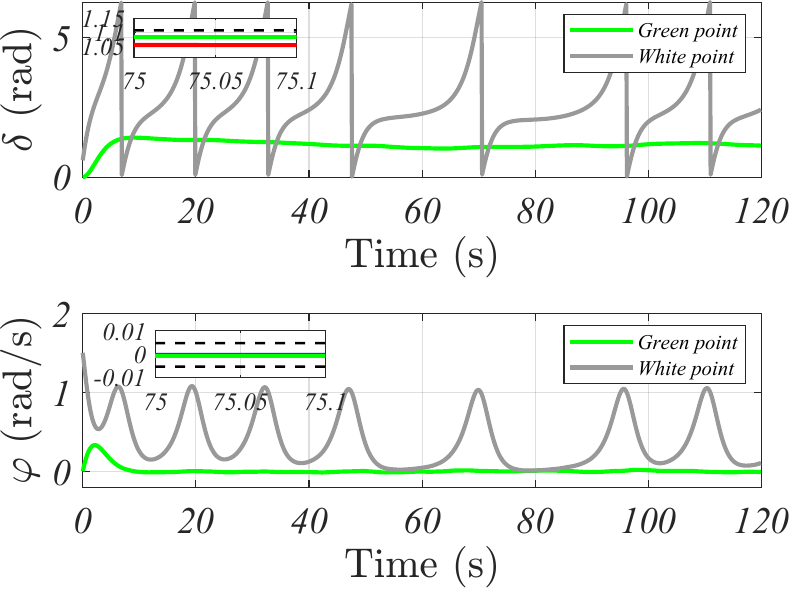} \\
     (c) & (d)
\end{tabular}
 \caption[]{%
    \protect\justifying \textbf{Stability basin of the synchronous solution as a function of the coupling parameter $K$ under variable wind conditions.}
    Panels~(a)--(b) report the case $K=4300~\mathrm{W}$, while panels~(c)--(d) correspond to $K=4500~\mathrm{W}$. Panels~(a) and~(c) display the basin of attraction of the synchronous solution in the plane of initial conditions $(\delta,\varphi)$. The green regions correspond to initial conditions leading to trajectories that converge to the synchronous state, whereas the white regions are associated with orbits that do not synchronize. Panels~(b) and~(d) show the time evolution of $\delta(t)$ and $\varphi(t)$ for two initial conditions: the point denoted by a circle lies in the stability basin, i.e., in the green region, while the point denoted by a square belongs to the white region. A clear difference between the two trajectories is observed: the former stabilizes around the reference solution, as indicated by the dashed lines in the insets, whereas the latter strongly deviates and exhibits persistent oscillations. The other parameters are fixed as follows: $D_{\mathrm{wind}}=0.5~\mathrm{rad/s}$, $H_{\mathrm{wind}}=40~\mathrm{kg.m^2}$, $V_{\mathrm{mean}}=3~\mathrm{m/s}$, $\epsilon_{\mathrm{OU}}=0.15$, and $\tau_{\mathrm{OU}}=60~\mathrm{s}$.
    }
    \label{fig:figure7}
\end{figure}
We now apply the above framework to analyze how the main system parameters affect the synchronization basin under stochastic wind fluctuations. Figure~\ref{fig:figure7} illustrates the influence of the coupling coefficient $K$ on the stability basin of a wind turbine generator subjected to stochastic wind fluctuations. {This analysis extends the previous basin approach to the variable-wind case, where synchronization is assessed using the error metrics defined above.} For a relatively low coupling value, $K = 4300~\mathrm{W}$, panels~(a,b) show that the green region is almost absent, indicating that only a very restricted set of initial conditions allows the system to remain stable and synchronized. This reflects the strong sensitivity to perturbations. The point marked by a circle belongs to the stable region, whereas the point marked by a square lies in the white region and therefore leads to a loss of synchronization. To highlight these behaviors, we illustrate in panel~(b) the temporal evolutions of $\delta(t)$, and $\varphi(t)$: the trajectory starting from the green point remains bounded within the prescribed error band and tends toward the synchronous regime, whereas the trajectory starting from the white point exhibits large oscillations and repeated deviations, indicating a loss of synchronization with the grid.

When the coupling strength is increased to $K = 4500~\mathrm{W}$, panel~(c,d) show a clear enlargement of the green region, meaning that the system becomes less sensitive to initial perturbations and that a larger set of initial conditions converges toward the synchronous state. {The temporal responses plotted in panel~(d) show that the green-point (e.g., the one identified by a circle) trajectory remains close to the reference orbit for both the phase and the frequency, while the white-point (e.g., the one identified by a square) trajectory still displays pronounced oscillations outside the admissible range.} These results demonstrate that increasing the coupling strength improves the ability of the system to recover synchronization after perturbations and enhances its robustness against perturbations and wind fluctuations.
\begin{figure}[htp!]
    \centering
\begin{tabular}{cc}
     \includegraphics[width=0.235\textwidth]{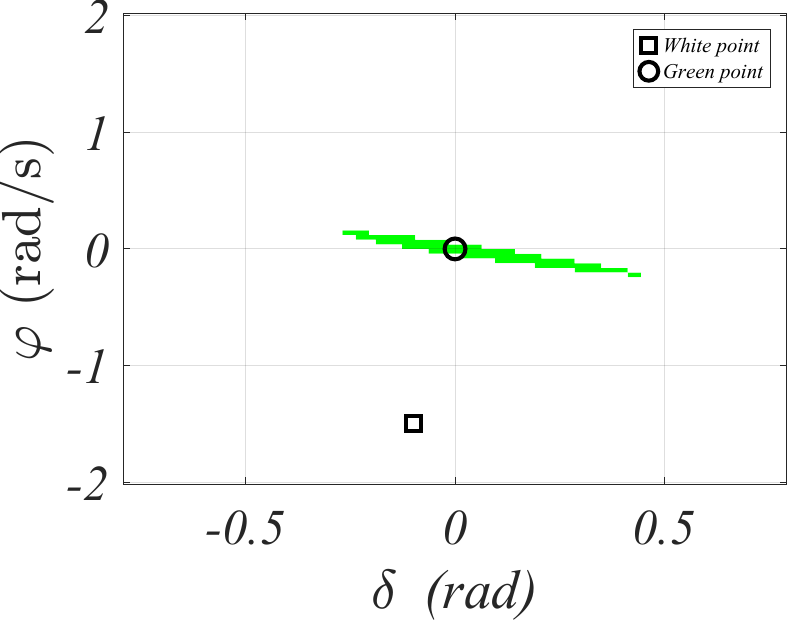}& 
     \includegraphics[width=0.235\textwidth]{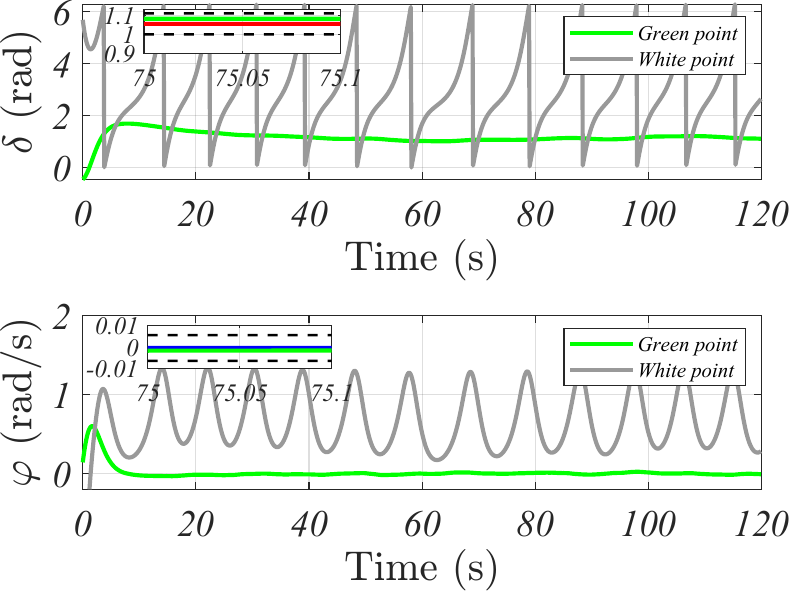}\\
     (a) & (b) \\
     \includegraphics[width=0.235\textwidth]{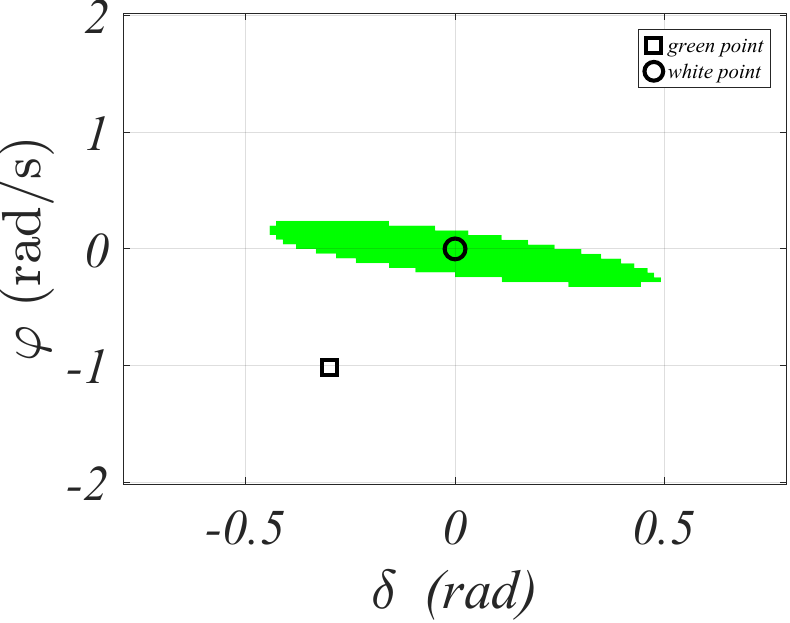}& 
     \includegraphics[width=0.235\textwidth]{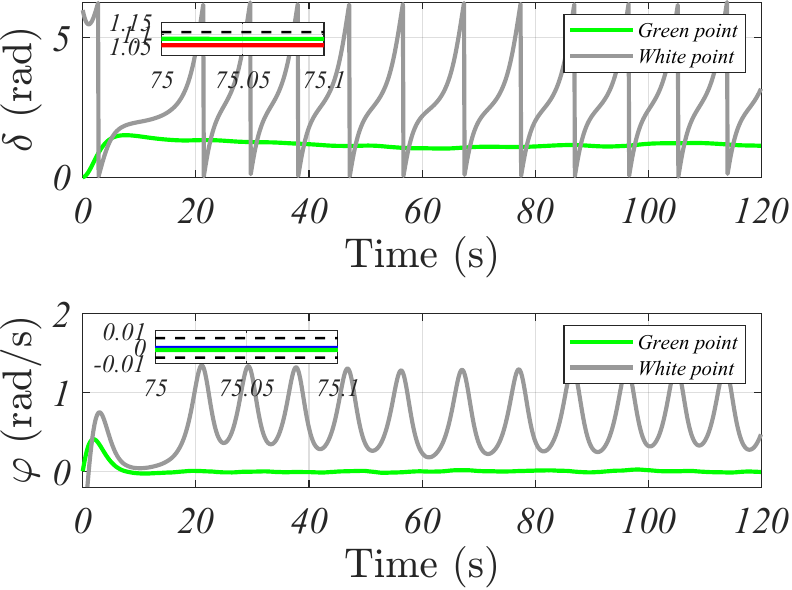} \\
     (c) & (d)
\end{tabular}
 \caption[]{%
    \protect\justifying 
  \textbf{Influence of the inertia parameter $H_{\mathrm{wind}}$ on the stability basin of the synchronous solution under variable wind conditions.} 
  Panels~(a)--(b) correspond to the case $H_{\mathrm{wind}}=10~\mathrm{kg.m^2}$, whereas panels~(c)--(d) correspond to $H_{\mathrm{wind}}=30~\mathrm{kg.m^2}$. Panels~(a) and~(c) show the basin of attraction of the synchronous solution in the plane of initial conditions $(\delta,\varphi)$. The green regions identify the initial conditions for which the resulting trajectories converge toward the synchronous state, while the white regions indicate initial conditions leading to a loss of synchronization. Panels~(b) and~(d) display the time evolution of $\delta(t)$ and $\varphi(t)$ for two representative initial conditions: the circle denotes a point located inside the stability basin, whereas the square denotes a point located in the white region. The corresponding trajectories exhibit markedly different behaviors: the trajectory starting from the circle remains close to the reference solution and stabilizes around it, as shown by the dashed lines in the insets, while the trajectory starting from the square deviates significantly and displays persistent oscillations. The other parameters are fixed at $D_{\mathrm{wind}}=0.5~\mathrm{rad/s}$, $V_{\mathrm{mean}}=3~\mathrm{m/s}$, $\epsilon_{\mathrm{OU}}=0.15$, $\tau_{\mathrm{OU}}=60~\mathrm{s}$, and $K=4500~\mathrm{W}$.
  }
    \label{fig:figure8}
\end{figure}
{Let us now investigate the effect of the inertia parameter on the synchronization basin when the wind turbine is subjected to wind fluctuations. Figure~\ref{fig:figure8} compares the system response for two values of the inertia $H_{\mathrm{wind}}$, while keeping the other parameters fixed.
}
For a relatively small inertia, $H_{\mathrm{wind}}=10~\mathrm{kg.m^2}$, panels~(a) and~(b) show that the synchronization basin is very narrow. Only a limited set of initial perturbations, represented by the green region, leads to trajectories that remain close to the reference solution and converge toward the synchronous state. The temporal responses in panel~(b) confirm this behavior: the trajectory starting from the green point (denoted by a circle) remains within the prescribed error band, whereas the trajectory starting from the white point (see the point denoted by a square) exhibits large oscillations and deviates from the synchronous regime. This scenario reflects rigidity against wind fluctuations and disturbances, making the stability of the system less robust to potential perturbations. When the inertia is increased to $H_{\mathrm{wind}}=30~\mathrm{kg.m^2}$, panels~(c) and~(d) show a clear enlargement of the synchronization basin of attraction. This indicates that a larger set of initial perturbations can be absorbed by the system without losing synchronization. The corresponding time evolutions in panel~(d) show that the trajectory associated with the green point stabilizes around the reference solution, while the trajectory associated with the white point still undergoes pronounced oscillations and does not synchronize.
{These results show that increasing the inertia improves the robustness of the wind turbine generator under variable wind conditions and which follow a stochastic process. Therefore, a larger inertia allows the system to better absorb perturbations and wind-induced fluctuations, thereby increasing the set of initial conditions leading to a stable synchronous state.}
\begin{figure}[htp!]
    \centering
\begin{tabular}{cc}
     \includegraphics[width=0.235\textwidth]{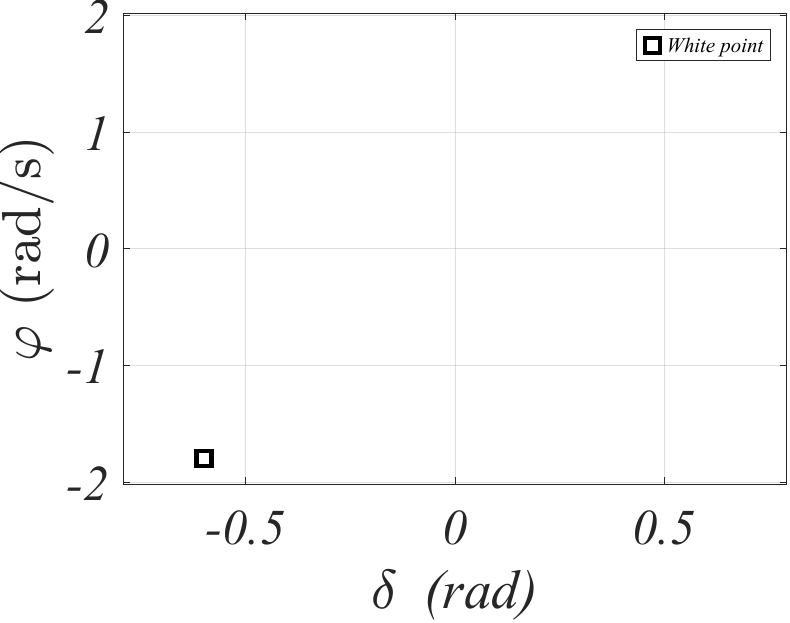}& 
     \includegraphics[width=0.235\textwidth]{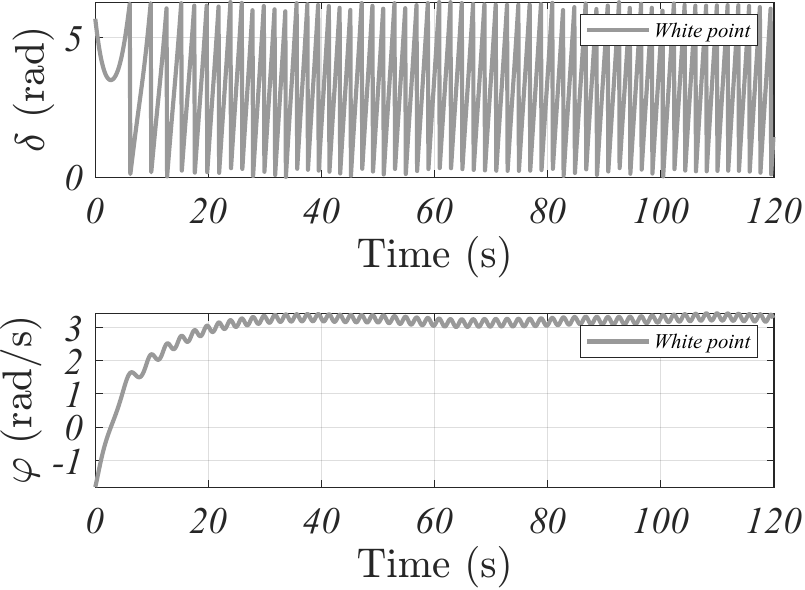}\\
     (a) & (b) \\
     \includegraphics[width=0.235\textwidth]{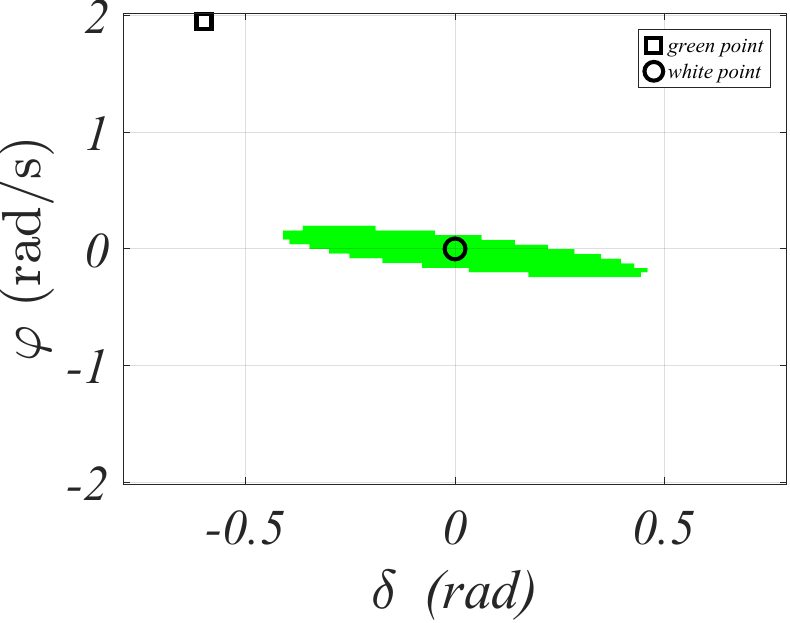}& 
     \includegraphics[width=0.235\textwidth]{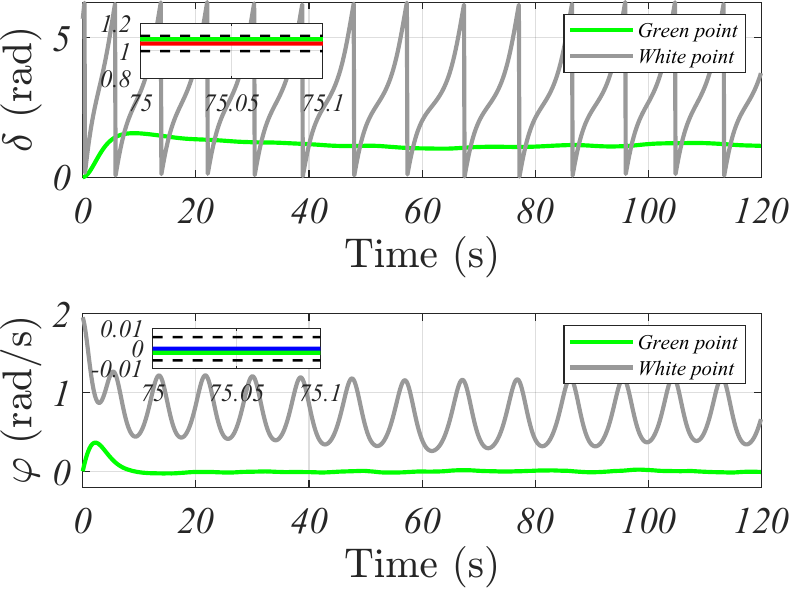} \\
     (c) & (d)
\end{tabular}
 \caption[]{%
    \protect\justifying \textbf{Effect of the damping coefficient $D_{\mathrm{wind}}$ on the stability basin of the synchronous solution under variable wind conditions.} Panels~(a)--(b) report the case of low damping with $D_{\mathrm{wind}}=0.1~\mathrm{rad/s}$, whereas panels~(c)--(d) correspond to relatively high damping with $D_{\mathrm{wind}}=0.4~\mathrm{rad/s}$. Panels~(a) and~(c) show the basin of attraction of the synchronous solution in the plane of initial conditions $(\delta,\varphi)$. The green regions identify the initial conditions for which the resulting trajectories converge toward the synchronous state, while the white regions indicate initial conditions leading to a loss of synchronization. Panel~(b) shows the time evolution of $\delta(t)$ and $\varphi(t)$ for one representative initial condition located in the white region and marked by a square. Panel~(d) presents the time evolution of the orbits for two representative initial conditions: the circle denotes a point inside the stability basin, whereas the square denotes a point located in the white region. The corresponding trajectories display a very different behaviors: the trajectory starting from the circle remains close to the reference solution and stabilizes around it, as shown by the dashed lines in the insets, while the trajectory starting from the square deviates strongly and exhibits persistent oscillations. The other parameters are fixed at $K=4500~\mathrm{W}$, $H_{\mathrm{wind}}=40~\mathrm{kg.m^2}$, $V_{\mathrm{mean}}=3~\mathrm{m/s}$, $\epsilon_{\mathrm{OU}}=0.15$, and $\tau_{\mathrm{OU}}=60~\mathrm{s}$.
    }
    \label{fig:figure9}
\end{figure}
{Let us examine the role of the damping coefficient in the synchronization properties of the wind turbine generator under variable wind conditions. According to the analysis done in the previous section for constant power $P$, this parameter is expected to play a key role in the dissipation of perturbation-induced energy and, therefore, in the ability of the system to recover synchronism after wind fluctuations. Figure~\ref{fig:figure9} compares the synchronization basin and the corresponding temporal responses for two values of $D_{\mathrm{wind}}$. For a very low damping value, $D_{\mathrm{wind}}=0.1~\mathrm{rad/s}$, panels~(a) and~(b) show that the green region is absent, indicating that no tested initial perturbation leads to stable synchronization. The selected trajectory, marked by a square and located in the white region (unstable domain), exhibits large and persistent oscillations in  $\delta(t)$ and $\varphi(t)$ outside the allowable error band. This behavior shows that weak damping does not allow the system to dissipate perturbation-induced energy efficiently, leading to a loss of synchronization with the grid. When the damping value is increased to $D_{\mathrm{wind}}=0.4~\mathrm{rad/s}$, panels~(c) and~(d) reveal the appearance of a well-defined green region. Initial conditions inside this gree region lead to trajectories that remain close to the reference solution and recover the synchronous state, as illustrated by the trajectory starting from the circle. In contrast, the trajectory starting from the square (see panel~(c)), located in the white region, deviates strongly from the reference solution and exhibits strong oscillations (see panel~(d)).
}
These results demonstrate that increasing the damping coefficient enhances the dissipation of electromechanical oscillations, promoting the return to the synchronous state and strengthening the wind generator's ability to maintain synchronization with the grid despite perturbations and wind speed fluctuations.

\begin{figure}[htp!]
    \centering
\begin{tabular}{cc}
     \includegraphics[width=0.235\textwidth]{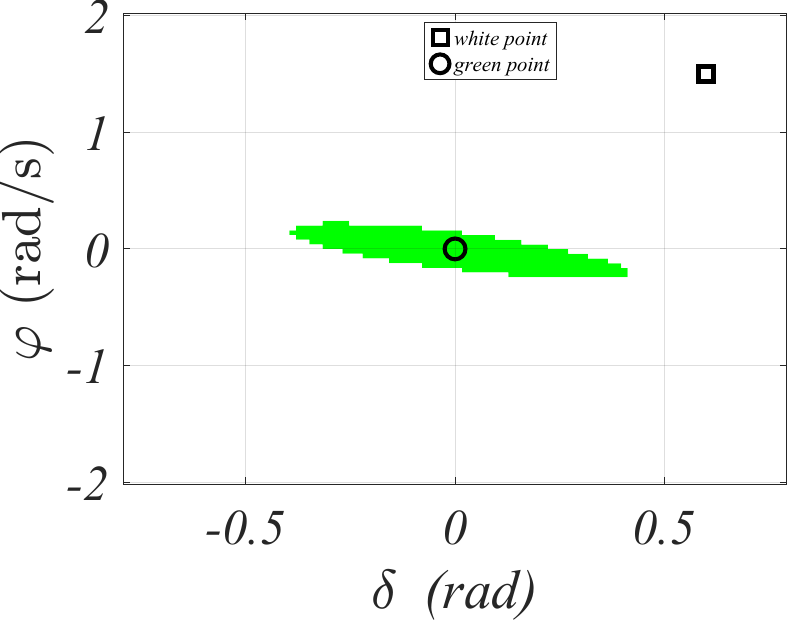}& 
     \includegraphics[width=0.235\textwidth]{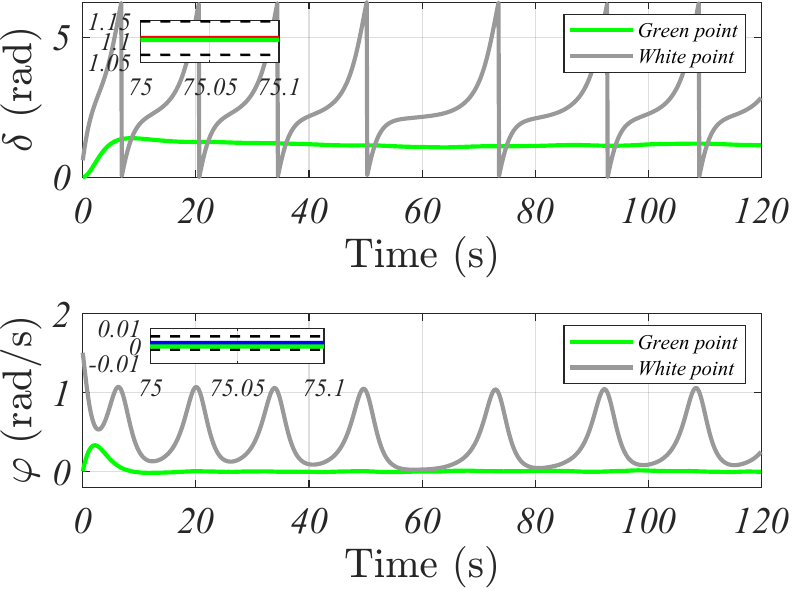}\\
     (a) & (b) \\
     \includegraphics[width=0.235\textwidth]{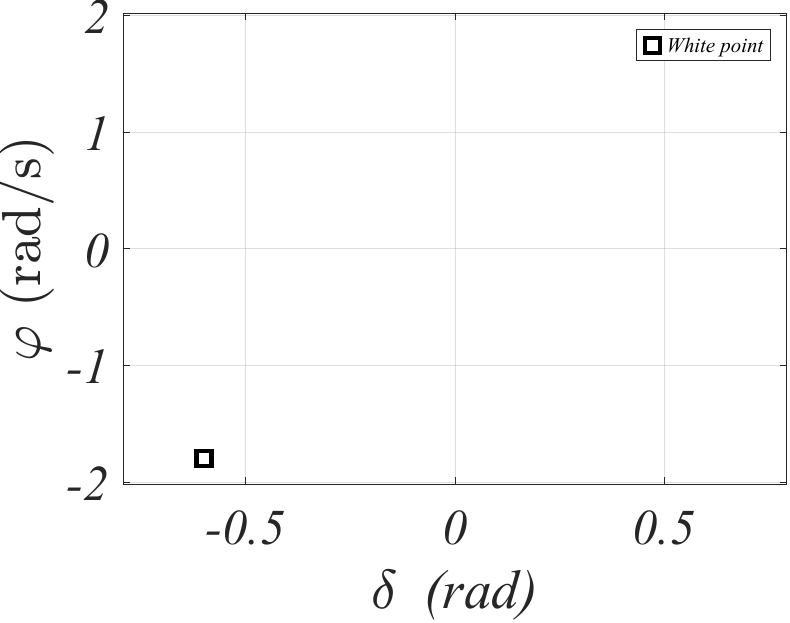}& 
     \includegraphics[width=0.235\textwidth]{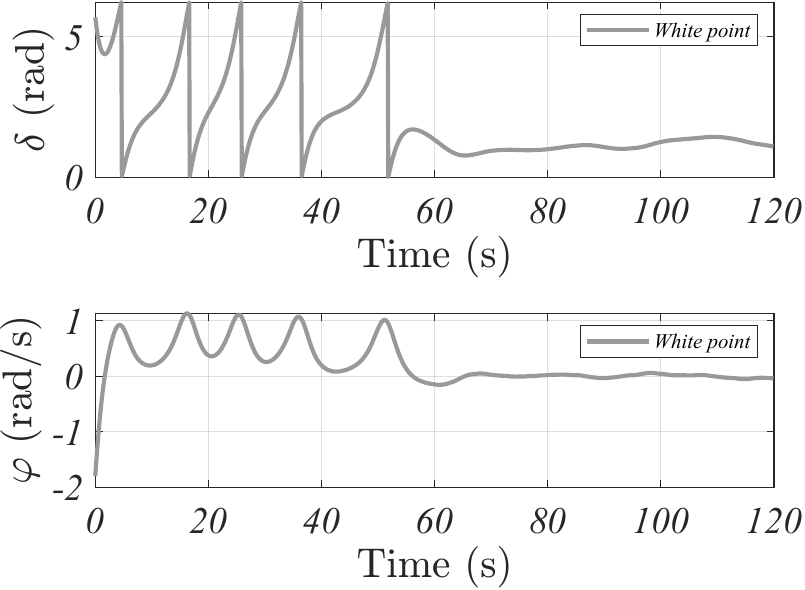} \\
     (c) & (d)
\end{tabular}
 \caption[]{%
    \protect\justifying \textbf{Influence of the wind fluctuation amplitude $\epsilon_{\mathrm{OU}}$ on the stability basin.} Panels~(a)--(b) correspond to $\epsilon_{\mathrm{OU}}=0.1$, whereas panels~(c)--(d) correspond to $\epsilon_{\mathrm{OU}}=0.5$. Panels~(a) and~(c) show the stability basin in the plane of initial conditions $(\delta,\varphi)$. The green regions denote initial conditions whose trajectories converge toward the synchronous state, while the white regions correspond to a loss of synchronization. In Panel~(b), we display the time evolution of the orbits for two representative initial conditions: the circle lies inside the green region, and the square located within the white region. The former remains close to the reference solution, indicated by the dashed lines in the insets, while the latter deviates strongly and exhibits persistent oscillations. Panel~(d) shows the response for a representative point in the white region, where the trajectory strongly deviates from the reference solution, confirming the loss of synchronization at high fluctuation intensity. The other parameters are fixed at $K=4500~\mathrm{W}$, $H_{\mathrm{wind}}=40~\mathrm{kg\,m^2}$, $V_{\mathrm{mean}}=3~\mathrm{m/s}$, $\tau_{\mathrm{OU}}=60~\mathrm{s}$, and $D_{\mathrm{wind}}=0.5~\mathrm{rad/s}$.}
    \label{fig:figure10}
\end{figure}

{After analyzing the effects of the coupling strength, inertia, and damping, let us now focus on the influence of the stochastic process introduced to model the wind fluctuation. In particular, we investigate how the amplitude of wind fluctuations, controlled by the parameter $\epsilon_{\mathrm{OU}}$, affects the ability of the wind turbine generator to maintain synchronization under variable wind conditions. Therefore, we illustrate in Figure~\ref{fig:figure10} the influence of the wind fluctuation amplitude $\epsilon_{\mathrm{OU}}$ on the stability basin of the system. For a relatively low fluctuation amplitude, $\epsilon_{\mathrm{OU}}=0.1$, panel~(a) shows that the green region (stable domain) extends over a wide portion of the initial perturbation plane $(\delta,\varphi)$. This indicates that for a large number of initial perturbations the system is still able to recover the synchronous state. Panel~(b) shows the corresponding temporal responses for a trajectory starting from the point marked by a circle, located inside the green region. This trajectory remains within the prescribed error band for both the phase and the frequency converges toward the synchronous state. In contrast, the trajectory starting from the point marked by a square, located in the white region, progressively deviates from the reference solution and indicating a loss of synchronization with the grid.}
{When the amplitude of the fluctuation is increased to $\epsilon_{\mathrm{OU}}=0.5$, panel~(c) shows that the stability basin is completely empty. In this case, none of the tested initial conditions leads to a stable synchronous state. This is further confirmed by the time evolutions in panel~(d), where the phase trajectories rapidly diverge and the frequency remains oscillatory throughout the simulation, outside the admissible error band. These results highlight the strongly destabilizing effect of high amplitude wind fluctuations. As $\epsilon_{\mathrm{OU}}$ increases, the stochastic variations of the wind power inject stronger dynamical perturbations into the system, reducing its ability to dissipate fluctuations and return to the synchronous state. Therefore, the amplitude of wind fluctuations plays a crucial role in determining the robustness of wind turbine generators and must be carefully considered when assessing synchronization stability in wind power networks.}
\begin{figure}[htp!]
    \centering
\begin{tabular}{cc}
     \includegraphics[width=0.235\textwidth]{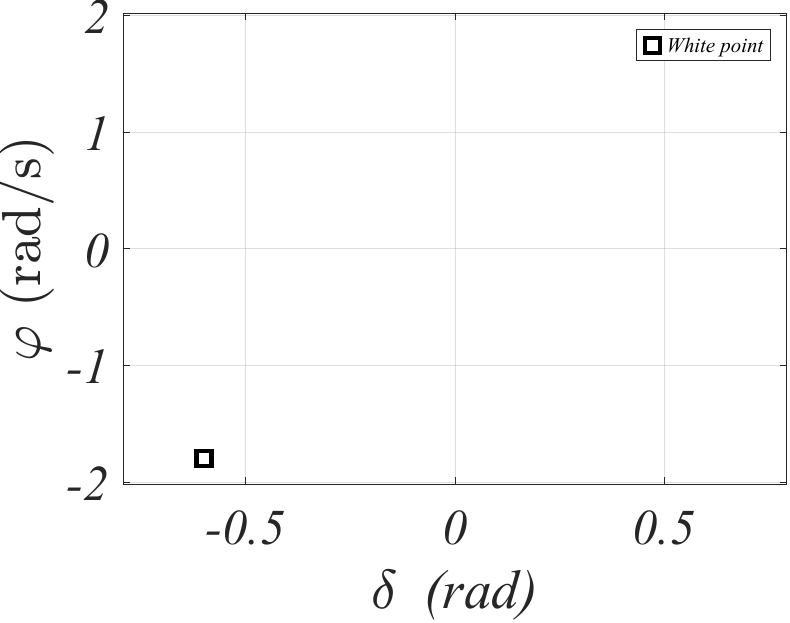}& 
     \includegraphics[width=0.235\textwidth]{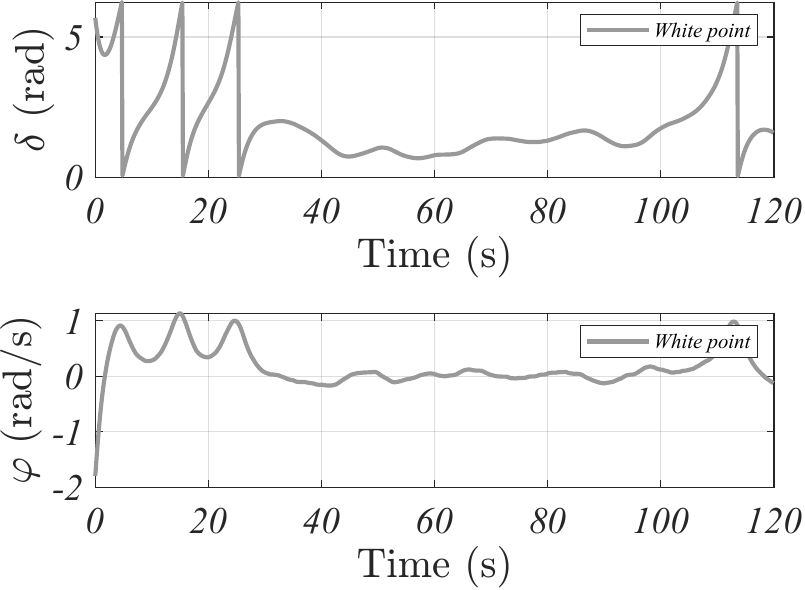}\\
     (a) & (b) \\
     \includegraphics[width=0.235\textwidth]{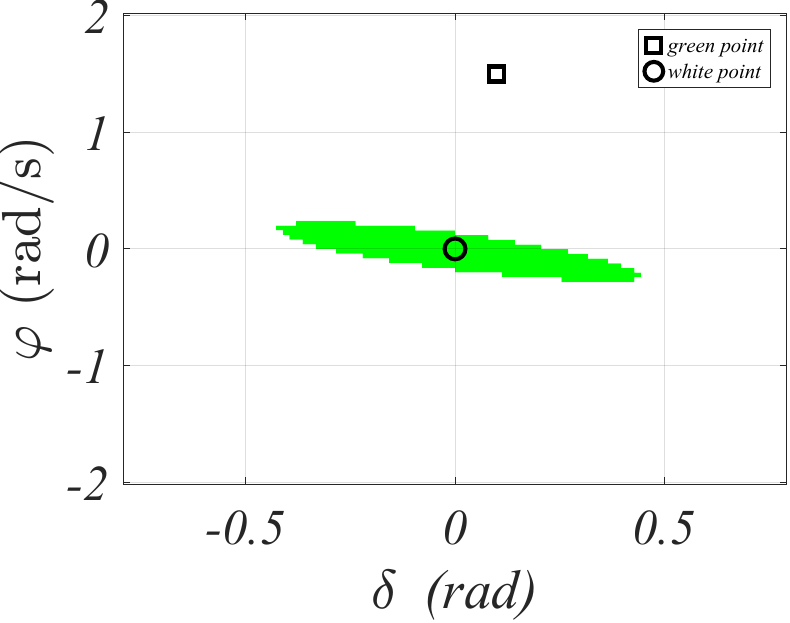}& 
     \includegraphics[width=0.235\textwidth]{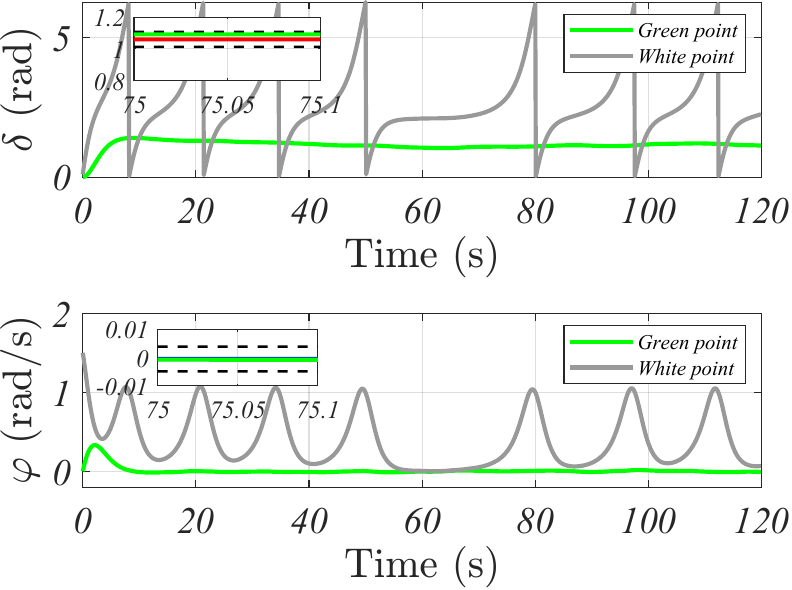} \\
     (c) & (d)
\end{tabular}
 \caption[]{%
    \protect\justifying \textbf{Effects of the correlation time $\tau_{\mathrm{OU}}$ on the stability basin under variable wind conditions.} Panels~(a)--(b) correspond to $\tau_{\mathrm{OU}}=5~\mathrm{s}$, whereas panels~(c)--(d) correspond to $\tau_{\mathrm{OU}}=70~\mathrm{s}$. Panels~(a) and~(c) show the stability basin in the plane of initial conditions $(\delta,\varphi)$: green regions indicate convergence to the synchronous state, while white regions indicate loss of synchronization. Panel~(b) presents the time evolution of $\delta(t)$ and $\varphi(t)$ for a representative initial condition marked by a square and located in the white region. Panel~(d) shows the time evolution of two representative orbits: the one starting from the circle, located in the green region, remains close to the reference solution, as indicated by the dashed lines in the insets, whereas the one starting from the square, located in the white region, deviates strongly and exhibits persistent oscillations. The other parameters are fixed at $D_{\mathrm{wind}}=0.5~\mathrm{rad/s}$, $K=4500~\mathrm{W}$, $H_{\mathrm{wind}}=40~\mathrm{kg\,m^2}$, $V_{\mathrm{mean}}=3~\mathrm{m/s}$, and $\epsilon_{\mathrm{OU}}=0.15$.}
    \label{fig:figure11}
\end{figure}
{Let us now focus on investigating the effect of the correlation time $\tau_{\mathrm{OU}}$ of wind fluctuations, which characterizes how rapidly the stochastic wind variations evolve in time. Figure~\ref{fig:figure11} illustrates and compares the stability basin and the corresponding temporal responses for short and long correlation times. For a short correlation time, $\tau_{\mathrm{OU}}=5~\mathrm{s}$, panel~(a) shows that the stability basin completely vanishes, indicating that none of the initial perturbations tested leads to a synchronous state. The corresponding temporal responses illustrated in panel~(b) confirm this behavior: the phase trajectory diverges, while the frequency exhibits oscillations outside the admissible error band. Physically, this loss of synchronization reflects the destabilizing effect of the fast variations of wind fluctuations, which act as high-frequency noise and strongly disturb the generator dynamics. When the correlation time increases to $\tau_{\mathrm{OU}}=70~\mathrm{s}$, panel~(c) shows the existence of a green domain, which is a characteristic of stability. In this case, a larger set of initial conditions allows the system to recover the synchronous state. The time responses in panel~(d), obtained from one initial condition located in the green region and another in the white region, support this observation. The trajectory starting from the green point remains within the error band for both phase and frequency and converges toward the synchronous state (stable), whereas the trajectory starting from the white point progressively deviates in phase and maintains frequency oscillations outside the admissible range (unstable).
} These results show that the temporal correlation of wind plays a determining role in system stability: fluctuations correlated over longer time scales act more gradually and are less destabilizing than rapid, highly noisy variations, which favors the maintenance of generator synchronization.

\section{Conclusion}
\label{sec:conc}
In this work we studied the synchronization and stability of wind power networks, with particularly attention to the case of significant disturbances that could destabilize the system. Systems composed of multiple interconnected wind generators, exhibit complex dynamics due to the natural and rapid variations in wind speed, as well as the coupled interactions between nodes. Understanding how these variations influence dynamic coherence and the network ability to remain synchronized is essential for ensuring the reliability and safety of electrical systems integrating intermittent renewable sources. The main objective of this study was twofold. On one hand, it aimed to quantify the impact of rapid wind speed fluctuations on network synchronization by analyzing the time responses of the generators to these variations. On the other hand, it sought to evaluate the dynamic stability of the network by considering key parameters such as generator inertia, coupling strength between nodes, rate of wind speed variation, correlation time, and damping, all of which play a major role in the system resilience. To achieve these objectives, we performed numerical simulations by using second order Kuramoto model with inertia, and by incorporating the interactions between generators and the effects of temporal perturbations in wind speed. {Since wind speed fluctuations induce variations in the generated wind power, they may strongly affect the system dynamics and, in some cases, lead to instability of the whole system.}
The simulations allow tracking the evolution of phases, angular velocities, and the order parameter $R$, which indicates the overall coherence of the network. The results obtained highlight that system synchronization and stability are highly sensitive to wind speed variations, but can be significantly improved by high inertia, strong coupling strength, and high damping. These conditions enhance the generators ability to quickly return to a synchronous state after a disturbance, thereby limiting the risks of instability or frequency deviation within the network.

\bibliographystyle{apsrev4-2}
\bibliography{sample}

@article{filatrella2008analysis,
  title={Analysis of a power grid using a Kuramoto-like model},
  author={Filatrella, Giovanni and Nielsen, Arne Hejde and Pedersen, Niels Falsig},
  journal={The European Physical Journal B},
  volume={61},
  pages={485--491},
  year={2008},
  publisher={Springer}
}

@article{Taher2019,
  author={Halgurd Taher and Simona Olmi and Eckehard Sch{\"o}ll},
  title={Enhancing power grid synchronization and stability through time-delayed feedback control},
  journal={Physical Review E},
  volume={100},
  number={6},
  pages={062306},
  year={2019},
  doi={10.1103/PhysRevE.100.062306},
  url={https://link.aps.org/doi/10.1103/PhysRevE.100.062306}
}

@article{Menck2014,
  author={Peter J. Menck and Jobst Heitzig and J{\"u}rgen Kurths and Hans Joachim Schellnhuber},
  title={How dead ends undermine power grid stability},
  journal={Nature Communications},
  volume={5},
  pages={3969},
  year={2014},
  doi={10.1038/ncomms4969},
  url={https://www.nature.com/articles/ncomms4969}
}

@article{luna2012optimal,
  title={Optimal sizing of renewable hybrids energy systems: A review of methodologies},
  author={Luna-Rubio, Ricardo and Trejo-Perea, Mario and Vargas-V{\'a}zquez, D and R{\'\i}os-Moreno, GJ},
  journal={Solar energy},
  volume={86},
  number={4},
  pages={1077--1088},
  year={2012},
  publisher={Elsevier}
}

@article{justus1978methods,
  title={Methods for estimating wind speed frequency distributions},
  author={Justus, CG and Hargraves, WR and Mikhail, Amir and Graber, Denise},
  journal={Journal of Applied Meteorology and Climatology},
  volume={17},
  number={3},
  pages={350--353},
  year={1978}
}

@article{takle1978characteristics,
  title={Characteristics of wind and wind energy in Iowa},
  author={Takle, ES and Brown, JM and Davis, WM},
  journal={Iowa State J. Res},
  volume={52},
  pages={313--339},
  year={1978}
}

@article{chauhan2014review,
  title={A review on Integrated Renewable Energy System based power generation for stand-alone applications: Configurations, storage options, sizing methodologies and control},
  author={Chauhan, Anurag and Saini, RP},
  journal={Renewable and Sustainable Energy Reviews},
  volume={38},
  pages={99--120},
  year={2014},
  publisher={Elsevier}
}

@article{black2006value,
  title={Value of storage in providing balancing services for electricity generation systems with high wind penetration},
  author={Black, Mary and Strbac, Goran},
  journal={Journal of power sources},
  volume={162},
  number={2},
  pages={949--953},
  year={2006},
  publisher={Elsevier}
}

@article{anagnostopoulos2007pumping,
  title={Pumping station design for a pumped-storage wind-hydro power plant},
  author={Anagnostopoulos, John S and Papantonis, Dimitris E},
  journal={Energy Conversion and Management},
  volume={48},
  number={11},
  pages={3009--3017},
  year={2007},
  publisher={Elsevier}
}

@article{carter1951analyzing,
  title={Analyzing winds for frequency and duration},
  author={Carter, JH and Gosline, CA and Hewson, EW and Landsberg, H and Barad, ML and Brier, GW and Hemeon, WCL and Lowry, PH and Mazzarella, DA and Smith, ME and others},
  journal={On Atmospheric Pollution: A Group of Contributions},
  pages={42--49},
  year={1951},
  publisher={Springer}
}

@article{bhandari2014novel,
  title={A novel off-grid hybrid power system comprised of solar photovoltaic, wind, and hydro energy sources},
  author={Bhandari, Binayak and Lee, Kyung-Tae and Lee, Caroline Sunyong and Song, Chul-Ki and Maskey, Ramesh K and Ahn, Sung-Hoon},
  journal={Applied Energy},
  volume={133},
  pages={236--242},
  year={2014},
  publisher={Elsevier}
}

@book{efstathios2012alternative,
  title={Alternative Energy Sources},
  author={Efstathios, E},
  year={2012},
  publisher={Springer}
}

@article{batista2021secondary,
  title={Secondary frequency control stabilizing voltage dynamics},
  author={Batista Tchawou Tchuisseu, Eder and Ndogmo, Eric-Donald and Proch{\'a}zka, Pavel and Woafo, Paul and Colet, Pere and Sch{\"a}fer, Benjamin},
  journal={arXiv e-prints},
  pages={arXiv--2109},
  year={2021}
}

@article{dorfler2014synchronization,
  title={Synchronization in complex networks of phase oscillators: A survey},
  author={D{\"o}rfler, Florian and Bullo, Francesco},
  journal={Automatica},
  volume={50},
  number={6},
  pages={1539--1564},
  year={2014},
  publisher={Elsevier}
}

@article{schafer2018non,
  title={Non-Gaussian power grid frequency fluctuations characterized by L{\'e}vy-stable laws and superstatistics},
  author={Sch{\"a}fer, Benjamin and Beck, Christian and Aihara, Kazuyuki and Witthaut, Dirk and Timme, Marc},
  journal={Nature Energy},
  volume={3},
  number={2},
  pages={119--126},
  year={2018},
  publisher={Nature Publishing Group UK London}
}

@article{schafer2015decentral,
  title={Decentral smart grid control},
  author={Sch{\"a}fer, Benjamin and Matthiae, Moritz and Timme, Marc and Witthaut, Dirk},
  journal={New journal of physics},
  volume={17},
  number={1},
  pages={015002},
  year={2015},
  publisher={IOP Publishing}
}

@article{menck2013basin,
  title={How basin stability complements the linear-stability paradigm},
  author={Menck, Peter J and Heitzig, Jobst and Marwan, Norbert and Kurths, J{\"u}rgen},
  journal={Nature physics},
  volume={9},
  number={2},
  pages={89--92},
  year={2013},
  publisher={Nature Publishing Group UK London}
}

@article{handbook2004union,
  title={Union for the Co-ordination of Transmission of Electricity},
  author={Handbook, UCTE Operation},
  journal={UCTE, 20th of July},
  year={2004}
}

@article{central2012report,
  title={Report on the grid disturbance on 30th July 2012 and grid disturbance on 31st July 2012},
  author={Central Electricity Regulatory Commission and others},
  journal={URL http://www. cercind. gov. in/2012/orders/final\_report\_grid\_disturbance. pdf},
  volume={450},
  year={2012}
}

@article{motter2002cascade,
  title={Cascade-based attacks on complex networks},
  author={Motter, Adilson E and Lai, Ying-Cheng},
  journal={Physical Review E},
  volume={66},
  number={6},
  pages={065102},
  year={2002},
  publisher={APS}
}

@article{schafer2018dynamically,
  title={Dynamically induced cascading failures in power grids},
  author={Sch{\"a}fer, Benjamin and Witthaut, Dirk and Timme, Marc and Latora, Vito},
  journal={Nature communications},
  volume={9},
  number={1},
  pages={1975},
  year={2018},
  publisher={Nature Publishing Group UK London}
}

@book{chiang2011direct,
  title={Direct methods for stability analysis of electric power systems: theoretical foundation, BCU methodologies, and applications},
  author={Chiang, Hsiao-Dong},
  year={2011},
  publisher={John Wiley \& Sons}
}

@article{tyloo2019noise,
  title={Noise-induced desynchronization and stochastic escape from equilibrium in complex networks},
  author={Tyloo, Melvyn and Delabays, Robin and Jacquod, Ph},
  journal={Physical Review E},
  volume={99},
  number={6},
  pages={062213},
  year={2019},
  publisher={APS}
}

@book{heier2014grid,
  title={Grid integration of wind energy: onshore and offshore conversion systems},
  author={Heier, Siegfried},
  year={2014},
  publisher={John Wiley \& Sons}
}

@article{gonzalez2023unveiling,
  title={Unveiling Inertia Constants by Exploring Mass Distribution in Wind Turbine Blades and Review of the Drive Train Parameters},
  author={Gonzalez-Rodriguez, Angel Gaspar and Roldan-Fernandez, Juan Manuel and Nieto-Nieto, Luis Miguel},
  journal={Machines},
  volume={11},
  number={9},
  pages={908},
  year={2023},
  publisher={MDPI}
}

@article{martinez2023dynamical,
  title={Dynamical model for power grid frequency fluctuations: Application to islands with high penetration of wind generation},
  author={Mart{\'\i}nez-Barbeito, Mar{\'\i}a and Gomila, Dami{\`a} and Colet, Pere},
  journal={IEEE Transactions on Sustainable Energy},
  volume={14},
  number={3},
  pages={1436--1445},
  year={2023},
  publisher={IEEE}
}

@article{tchuisseu2017effects,
  title={Effects of dynamic-demand-control appliances on the power grid frequency},
  author={Tchuisseu, EB Tchawou and Gomila, Dami{\`a} and Brunner, Daniel and Colet, Pere},
  journal={Physical Review E},
  volume={96},
  number={2},
  pages={022302},
  year={2017},
  publisher={APS}
}

@article{pham2025application,
  title={Application of the Ornstein--Uhlenbeck process to generate stochastic vertical wind profiles},
  author={Pham, Khiem and Nguyen, Anh Tuan and Nguyen, Linh Ngoc and Van Do Thi, Thanh},
  journal={Journal of Aircraft},
  volume={62},
  number={4},
  pages={1067--1071},
  year={2025},
  publisher={American Institute of Aeronautics and Astronautics}
}

@article{jonsdottir2020stochastic,
  title={Stochastic modeling of tidal generation for transient stability analysis: A case study based on the all-island Irish transmission system},
  author={J{\'o}nsd{\'o}ttir, Gu{\dh}r{\'u}n Margr{\'e}t and Milano, Federico},
  journal={Electric Power Systems Research},
  volume={189},
  pages={106673},
  year={2020},
  publisher={Elsevier}
}

@article{maller2009ornstein,
  title={Ornstein--Uhlenbeck processes and extensions},
  author={Maller, Ross A and M{\"u}ller, Gernot and Szimayer, Alex},
  journal={Handbook of financial time series},
  pages={421--437},
  year={2009},
  publisher={Springer}
}

@article{tchuisseu2023secondary,
  title={Secondary frequency control stabilising voltage dynamics},
  author={Tchuisseu, Eder Batista Tchawou and Dongmo, Eric-Donald and Proch{\'a}zka, Pavel and Woafo, Paul and Colet, Pere and Sch{\"a}fer, Benjamin},
  journal={European Journal of Applied Mathematics},
  volume={34},
  number={3},
  pages={467--483},
  year={2023},
  publisher={Cambridge University Press}
}

@article{taher2019enhancing,
  title={Enhancing power grid synchronization and stability through time-delayed feedback control},
  author={Taher, Halgurd and Olmi, Simona and Sch{\"o}ll, Eckehard},
  journal={Physical Review E},
  volume={100},
  number={6},
  pages={062306},
  year={2019},
  publisher={APS}
}

@article{nguyen2020electric,
  title={Electric power grid resilience to cyber adversaries: State of the art},
  author={Nguyen, Tien and Wang, Shiyuan and Alhazmi, Mohannad and Nazemi, Mostafa and Estebsari, Abouzar and Dehghanian, Payman},
  journal={IEEE access},
  volume={8},
  pages={87592--87608},
  year={2020},
  publisher={IEEE}
}

@article{acebron2005kuramoto,
  title={The Kuramoto model: A simple paradigm for synchronization phenomena},
  author={Acebr{\'o}n, Juan A and Bonilla, Luis L and P{\'e}rez Vicente, Conrad J and Ritort, F{\'e}lix and Spigler, Renato},
  journal={Reviews of modern physics},
  volume={77},
  number={1},
  pages={137--185},
  year={2005},
  publisher={APS}
}

@article{hu2010parameter,
  title={Parameter estimation for fractional Ornstein--Uhlenbeck processes},
  author={Hu, Yaozhong and Nualart, David},
  journal={Statistics \& probability letters},
  volume={80},
  number={11-12},
  pages={1030--1038},
  year={2010},
  publisher={Elsevier}
}

@article{kavade2025variable,
  title={Variable pitching study of small scale vertical axis wind turbine},
  author={Kavade, Ramesh K and Jaiswal, Naresh and Bhone, Nitin P and Kothare, Chandrakant B and Urade, Ashish D and Kolhe, Ajay V and Haque Siddiqui, Md Irfanul and Chan, Choon Kit and Yong, Xu},
  journal={Energy Exploration \& Exploitation},
  volume={43},
  number={6},
  pages={2424--2445},
  year={2025},
  publisher={SAGE Publications Sage UK: London, England}
}

@article{gillespie1996exact,
  title={Exact numerical simulation of the Ornstein-Uhlenbeck process and its integral},
  author={Gillespie, Daniel T},
  journal={Physical review E},
  volume={54},
  number={2},
  pages={2084},
  year={1996},
  publisher={APS}
}

@article{ren2018analysis,
  title={The analysis of turbulence intensity based on wind speed data in onshore wind farms},
  author={Ren, Guorui and Liu, Jinfu and Wan, Jie and Li, Fei and Guo, Yufeng and Yu, Daren},
  journal={Renewable energy},
  volume={123},
  pages={756--766},
  year={2018},
  publisher={Elsevier}
}

@book{stiebler2008wind,
  title={Wind energy systems for electric power generation},
  author={Stiebler, Manfred},
  year={2008},
  publisher={Springer Science \& Business Media}
}

@misc{irm_climate,
  author       = {Royal Meteorological Institute of Belgium },
  title        = {Climatological Normals of Belgium},
  year         = {2020},
  url          = {https://www.meteo.be/en/belgium},
 }

\end{document}